\documentclass[aps,prl,showpacs,showkeys,amsmath,amssymb,12pt
]{article}
\pdfoutput=1

 \usepackage{multirow}

\usepackage{cite}
\usepackage{amsmath,bbm}
\usepackage{amssymb}
\usepackage{graphicx} 
\usepackage[latin1]{inputenc}
\usepackage{mathrsfs}
\usepackage{mathtools}
\usepackage[dvipsnames]{xcolor}
\usepackage[footnotesize]{caption}
\usepackage{xcolor}
\usepackage{hyperref}
\usepackage[hmarginratio=1:1,top=32mm,columnsep=20pt]{geometry}
\usepackage{paralist}
\usepackage{abstract}
\usepackage{upgreek}

\hypersetup{colorlinks,bookmarksopen,bookmarksnumbered,citecolor=blue,
linkcolor=black,pdfstartview=FitH,urlcolor=blue}

\newcommand{\parenbar}[1]{\overset{
            \raisebox{-0.15em}{\scalebox{.4}{\textbf{(}}}
            \raisebox{-0.3em}{{\hspace{.03em}--\hspace{.05em}}}
            \raisebox{-0.15em}{\scalebox{.4}{\textbf{)}}}} {#1}}

\newcommand{\GammaN}{\Gamma_{N\to\nu\gamma}}

\allowdisplaybreaks[4] 

\date{}
\begin{document}

\begin{titlepage}

\renewcommand{\thefootnote}{\alph{footnote}}
\title{\bf Explaining the MiniBooNE excess by a decaying sterile neutrino with mass in the 250~MeV range}

\author{Oliver Fischer, \'Alvaro Hern\'andez-Cabezudo, and Thomas Schwetz\\[3mm]
\small  
{\it Institut f\"ur Kernphysik, Karlsruhe Institute of Technology (KIT),}\\[-1mm]
\small  
{\it Hermann-von-Helmholtz-Platz 1, 76344 Eggenstein-Leopoldshafen, Germany}
}

\end{titlepage}

\maketitle

\begin{abstract}
The MiniBooNE collaboration has reported an excess of $460.5\pm 95.8$ electron-like events ($4.8\sigma$). We propose an explanation of these events in terms of a sterile neutrino decaying into a photon and a light neutrino. The sterile neutrino has a mass around 250~MeV and it is produced from kaon decays in the proton beam target via mixing with the muon or the electron in the range $10^{-11} \lesssim |U_{\ell 4}|^2 \lesssim 10^{-7}$ ($\ell = e,\mu$). The model can be tested by considering the time distribution of the events in MiniBooNE and by looking for single-photon events in running or upcoming neutrino experiments, in particular by the suite of liquid argon detectors in the short-baseline neutrino program at Fermilab.
\end{abstract}


\newpage
\tableofcontents

\hypersetup{linkcolor=blue}

\section{Introduction}
The MiniBooNE collaboration has published evidence for an excess of electron-like events of $381.2 \pm 85.2$ above their background expectation \cite{Aguilar-Arevalo:2018gpe}, confirming previous hints present in both, neutrino and anti-neutrino beam modes \cite{Aguilar-Arevalo:2013pmq}. The combined excess of $460.5\pm 95.8$ events corresponds to a significance of $4.8\sigma$.  The collaboration presents the results in the context of $\parenbar{\nu}_\mu\to\parenbar{\nu}_e$ neutrino oscillations, under the hypothesis of a sterile neutrino with a neutrino mass-squared difference $\Delta m^2$ of order 1~eV$^2$, motivated by a previous claim from LSND~\cite{Aguilar:2001ty}. 
The interpretation of the above mentioned results in terms of neutrino oscillations with an eV-scale sterile neutrino is in strong conflict with data on $\parenbar{\nu}_e$ and $\parenbar{\nu}_\mu$ neutrino disappearance at the $\Delta m^2 \sim 1$~eV$^2$ scale \cite{Dentler:2018sju, Gariazzo:2017fdh, Diaz:2019fwt}. This motivates to look for other new-physics explanations, beyond sterile neutrino oscillations. 

In this paper we propose a sterile neutrino in the 150 to 300~MeV mass range, which is produced in the beam target from kaon decay via mixing either with electron or muon neutrinos. Subsequently it decays inside the MiniBooNE detector into a photon and a light neutrino. Since the electromagnetic shower of a photon inside MiniBooNE cannot be distinguished from the one of an electron or positron the photon can explain the observed excess events. We study the energy and angular spectra and predict a specific time distribution of the events. In order to obtain a reasonable fit to the angular distribution, we are driven to heavy neutrino masses around 250~MeV, which can be produced by kaon decays in the beam target. For lighter neutrino decays, the signal is too much forward peaked, inconsistent with MiniBooNE data~\cite{Jordan:2018qiy}. Then the heavy neutrinos are only moderately relativistic and therefore our signal has a specific time structure, which provides a testable signature of our model~\cite{Ballett:2016opr}. The required parameters are consistent with all laboratory, astrophysics, and cosmology bounds. Current bounds and sensitivities of the upcoming short-baseline program at Fermilab for $N \to \gamma \nu$ with the heavy neutrino $N$ in the relevant mass range have been discussed in ref.~\cite{Ballett:2016opr}.
Our model differs from various previously discussed explanations of
the MiniBooNE and LSND anomalies based on the decay of a sterile
neutrino. In the explanations of
refs.~\cite{Gninenko:2009ks,Gninenko:2010pr} and
\cite{Bertuzzo:2018itn, Ballett:2018ynz} the heavy neutrino is
produced by $\nu_\mu$ scattering inside the detector and has to decay
with a very short lifetime into a photon or an $e^\pm$ pair,
respectively. The photon model from ref.~\cite{Gninenko:2010pr} is by
now excluded by searches for radiative neutrino decays from kaons by
the ISTRA+ experiment \cite{Duk:2011yv}, see also
\cite{Masip:2012ke, Magill:2018jla}. For other decay scenarios and related work see
refs.~\cite{Ma:1999im, PalomaresRuiz:2005vf, Dib:2011hc, 
  Arguelles:2018mtc, Diaz:2019fwt}.

The article is structured as follows. In section~\ref{sec:model} we
introduce the model and in section~\ref{sec:events} we describe the
calculation of the MiniBooNE signal, including the time, energy, and
angular event distributions. We present our $\chi^2$ fit to the data
in section~\ref{sec:fit}. The results are discussed in terms of the
model parameters in section~\ref{sec:results}, which includes also a
discussion of other constraints on the model and possible tests in
existing or upcoming experiments. In section~\ref{sec:conclusions} we
conclude. Details of the heavy neutrino flux calculation are given in
appendix~\ref{sec:flux}, in appendix~\ref{app:timing} we discuss the
impact of the timing cut on the MiniBooNE fit result.

\section{The model}\label{sec:model}

We consider one heavy Dirac neutrino $N$ with mass $m_N$ that mixes with the SM neutrinos, parameterized by the leptonic mixing matrix $U$. The sub-matrix $U_{\ell i}$ with $\ell = e,\mu,\tau$ and $i=1,2,3$ is approximately the PMNS matrix that gives rise to neutrino oscillations, and the matrix elements $U_{\ell 4}$ allow $N$ to interact with the weak currents and the lepton doublets of the Standard Model.
Focusing on the case $m_N = {\cal O}(100)$~MeV, we consider effective four-fermion interactions between the heavy neutrino and the SM particles, which are the mesons and leptons at this energy scale. Of particular importance is the following effective operator:
\begin{equation}
  {\cal O}_{\ell N q_u q_d} = U_{\ell 4} V_{q_u q_d}\,G_F \,
   \left[ \bar q_u \gamma^\mu(1-\gamma_5)q_d\right] \,
   \left[ \bar \ell \gamma_\mu(1-\gamma_5) N \right] + \text{h.c.}\,,
\label{eq:operator1}
\end{equation}
where $G_F$ is the Fermi constant, $q_u$ and $q_d$ are up-type and down-type quarks, respectively, $V$ is the CKM matrix, and $\ell$ is a charged lepton. 
Fixing the CKM matrix element to $V_{us}$, the operator in eq.~\eqref{eq:operator1} allows us to calculate the branching ratio of the kaon into a lepton $\ell=e,\mu$ and the neutrino $N$. For later use we define the following quantity:
\begin{align}
\rho_\ell(m_N) &\equiv
\frac{\text{Br}(K \to \ell N)}{\text{Br}(K \to \mu\nu)}  \nonumber\\
& = \frac{\text{Br}(K \to \ell \nu)}{\text{Br}(K \to \mu\nu)}\, |U_{\ell 4}|^2\, \frac{\left(x_N^2 + x_\ell^2- (x_N^2-x_\ell^2)^2 \right)\sqrt{(1-(x_N+x_\ell)^2)(1-(x_N-x_\ell)^2)}}{x_\ell^2(1 - x_\ell^2)^2}\,, \label{eq:rho}
\end{align}
which takes into account the mixing of the heavy neutrino and the kinematical factors related to the finite mass of the neutrino~\cite{Shrock:1980ct}. 
Here, $x_i = m_i/m_K$ and we use 
Br$(K\to\mu \nu) = 0.636$ and Br$(K\to e \nu) = 1.6 \times 10^{-5}$.
The factor $\rho_\ell(m_N)$ is normalized to the branching ratio of $K\to\mu\nu$, since we use the kaon induced $\parenbar{\nu}_\mu$ flux in MiniBooNE to derive the heavy neutrino flux in both cases, $K\to N\mu$ and $K\to N e$, see appendix~\ref{sec:flux}. 

In order to obtain the decay $N\to\nu\gamma$  into a light neutrino and a photon we introduce another effective operator to parameterize the possible interaction of $N$ with a photon and light neutrinos via its magnetic moment~\cite{Gninenko:2009ks, Aparici:2009fh}
\footnote{The operator in eq.~\eqref{eq:operator2} has been chosen as a specific example for a possible decay mechanism, which we use below to study the relevant phenomenology. Other operators (including dimension-6 operators) inducing $N\to\nu\gamma$ in the case of Majorana neutrinos have been considered e.g., in refs.~\cite{Duarte:2015iba,Butterworth:2019iff}.}:
\begin{equation}
{\cal O}_{N\nu\gamma} =  \frac{1}{\Lambda} \bar N \sigma^{\alpha\beta} \nu F_{\alpha \beta}\,,
\label{eq:operator2}
\end{equation}
with the electromagnetic field strength tensor $F_{\mu\nu} = \partial_\mu A_\nu - \partial_\nu A_\mu$, the anti-symmetric tensor $\sigma^{\mu\nu} = \gamma^\mu\gamma^\nu-\gamma^\nu\gamma^\mu$, and the unknown energy scale $\Lambda$. The operator ${\cal O}_{N\nu\gamma}$ could be created at the loop level, for instance, such that we expect $1/\Lambda$ to be a combination of an inverse mass, unknown coupling constants, and a typical loop suppression factor. 
The operator in eq.~\eqref{eq:operator2} allows $N$ to decay via the process $N\to \nu \gamma$, with the total width in the rest frame of $N$ given by 
\begin{equation}\label{eq:Gamma-rf}
\GammaN =  \frac{m_N^3}{4 \pi \Lambda^2} \approx
    1.2\times 10^{-16} \, {\rm MeV}
    \left( \frac{10^5 \,{\rm TeV}}{\Lambda}\right)^2
    \left( \frac{m_N}{250 \,{\rm MeV}}\right)^3 \,.
\end{equation}
To predict the energy and angular event spectra in MiniBooNE, we will need the differential decay rates with respect to the photon momentum $p_\gamma$ and the angle $\theta$ between the photon and $N$ momenta in the laboratory frame:
\begin{align}
\frac{d\Gamma_{N \to \nu \gamma}^{\rm lab}}{d p_\gamma} &= \frac{1}{4 \pi \Lambda^2}\frac{m_N^4}{E_N p_N }\,,
\label{eq:dGammadp} \\
\frac{d\Gamma_{N \to \nu \gamma}^{\rm lab}}{d \cos \theta} &= \frac{1}{8\pi \Lambda^2 E_N} \frac{m_N^6}{(E_N-p_N\cos\theta)^2}\,.
\label{eq:dGammadz}
\end{align}
The minimum value of $p_\gamma$ is in backward direction, $p_{\gamma, \rm min} = (E_N - p_N)/2$, and the maximum value in forward direction, $p_{\gamma,\rm max} = (E_N + p_N)/2$.

The phenomenology of the magnetic moment operator from
eq.~\eqref{eq:operator2} has been studied extensively in
ref.~\cite{Magill:2018jla}, see also
\cite{Ballett:2016opr,Coloma:2017ppo,Shoemaker:2018vii} for recent
considerations. In general this operator provides also a production
channel for the heavy neutrinos \cite{Masip:2012ke,
  Magill:2018jla}. Comparing with the results of
ref.~\cite{Magill:2018jla} we will see that for decay rates relevant
for our scenario, the production via mixing and weak boson mediated
kaon decay as described in relation to eq.~\eqref{eq:rho} will be the
dominant production mechanism.

The neutrino mixing parameters $U_{\ell 4}$ allow for various decay modes of $N$ into SM particles via weak boson exchange; depending on its mass into a number of leptons, or also into a lepton and one or more mesons, which have been computed e.g.\ in refs.~\cite{Atre:2009rg, Bondarenko:2018ptm}. In the mass range of interest to us, $m_\pi < m_N < m_K$, the dominant decay modes are $N\to \ell^\pm \pi^\mp$ and $N\to \nu\pi^0$. Using the results of ref.~\cite{Bondarenko:2018ptm} the decay rate can be estimated by 
\begin{align}
  \Gamma_\pi \equiv \Gamma_{N\to {\rm lept} \pi} &= \frac{G_F^2 f_h^2 m_N^3}{32\pi} |U_{\ell 4}|^2 g(m_\pi, m_{\rm lept}, m_N)\nonumber\\
  &\approx 3\times 10^{-13} \,{\rm MeV} \, |U_{\ell 4}|^2 
    \left( \frac{m_N}{250 \,{\rm MeV}}\right)^3 g(m_\pi, m_{\rm lept}, m_N)\,. \label{eq:GammaSM}
\end{align}
Here, $g(m_\pi, m_{\rm lept}, m_N)$ is a dimensionless kinematical function depending on the decay channel~\cite{Bondarenko:2018ptm}, "lept" indicates either a light neutrino or a charged lepton of flavour $\ell=e,\mu$, and $f_h \approx 130$~MeV is the pion decay constant. As we will see below, for large portions of the parameter space for $\GammaN$ and $U_{\ell 4}$ required to explain the MiniBooNE events, the decays $N\to {\rm lept} \pi$ will be sub-leading compared to $N\to\nu\gamma$.

Note that a decay width of the scale indicated in eq.~\eqref{eq:Gamma-rf} corresponds to lifetimes much shorter than milliseconds, and therefore our sterile neutrino decays well before Big Bang nucleo-synthesis and hence does not affect cosmology. Heavy neutrinos in the 100~MeV mass range are at the border of being relevant for supernova cooling arguments. Limits from supernova 1987A on heavy neutrino mixing are avoided in our scenario \cite{Dolgov:2000jw}, while the limits due to the magnetic moment operator derived in ref.~\cite{Magill:2018jla} will be relevant in part of the parameter space able to explain the MiniBooNE excess, see also \cite{Fuller:2009zz, Rembiasz:2018lok}.

\bigskip

To summarize, the relevant phenomenology of our model is determined by
three independent parameters, which we chose to be the heavy neutrino
mass: $m_N$, the mixing with the $e$ or $\mu$ flavour: $|U_{\ell
  4}|^2$, and the decay width into the photon: $\GammaN$. We will
present the parameter space where the MiniBooNE excess can be
explained in terms of those three parameters in
section~\ref{sec:results} below.

\section{The MiniBooNE excess events}\label{sec:events}

Our analysis proceeds as follows: first we construct the kaon flux at
the BNB from the given flux of the muon neutrinos. From the kaon flux
we derive the flux of the heavy neutrinos and work out its time
structure.  Then we calculate the energy and angular spectra of the
photon from the heavy neutrino decays inside the detector and inside
the time window defined by the MiniBooNE collaboration.

In order to calculate the flux of heavy neutrinos $\Phi_N(p_N)$ we
proceed as follows. We depart from the kaon contribution to the
$\parenbar{\nu}_\mu$ fluxes provided by the MiniBooNE collaboration
ref.~\cite{AguilarArevalo:2008yp}. Assuming that this flux is
dominated by the two-body decay $K\to\nu\mu$ we reconstruct the
initial kaon flux, from which in turn we can calculate the heavy
neutrino flux at MiniBooNE by taking into account the modified angular
acceptance of the detector due to the non-negligible effect of the
heavy neutrino mass on the angular distribution. Details of this
procedure are provided in appendix~\ref{sec:flux}. Note that the flux
$\Phi_N(p_N)$ obtained in this way depends on the mass of the heavy
neutrino, which we keep implicit to simplify notation.

\subsection{Time spectrum}\label{sec:timing}
\begin{figure}
    \centering
    \includegraphics[width=0.4\textwidth]{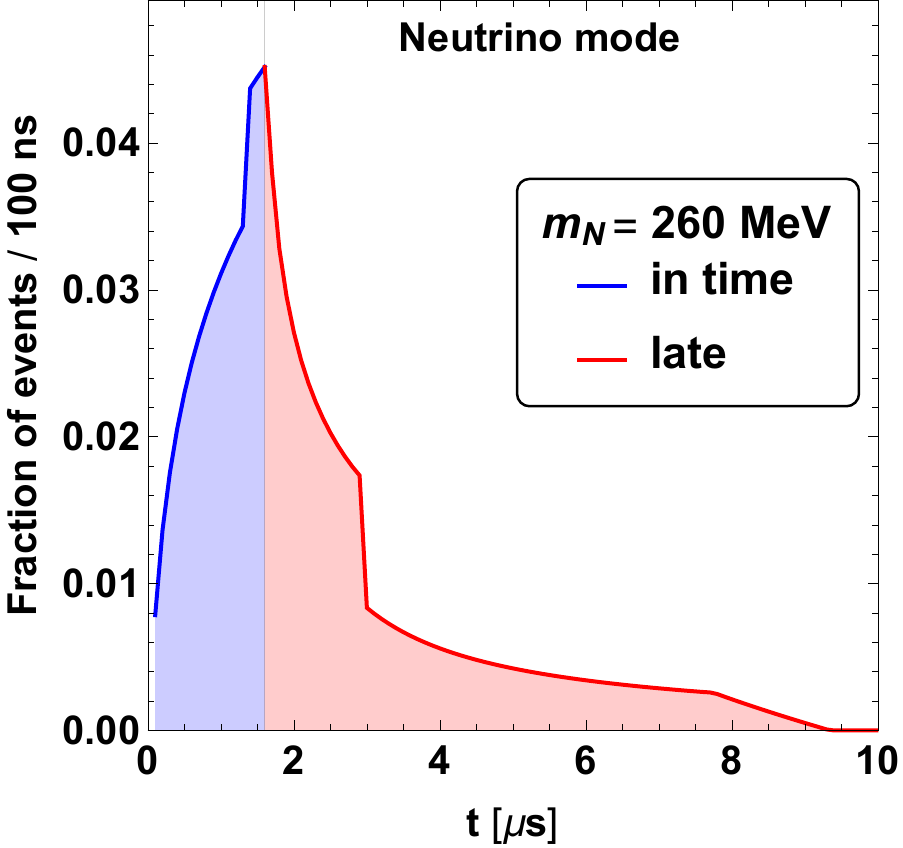}\qquad
    \includegraphics[width=0.4\textwidth]{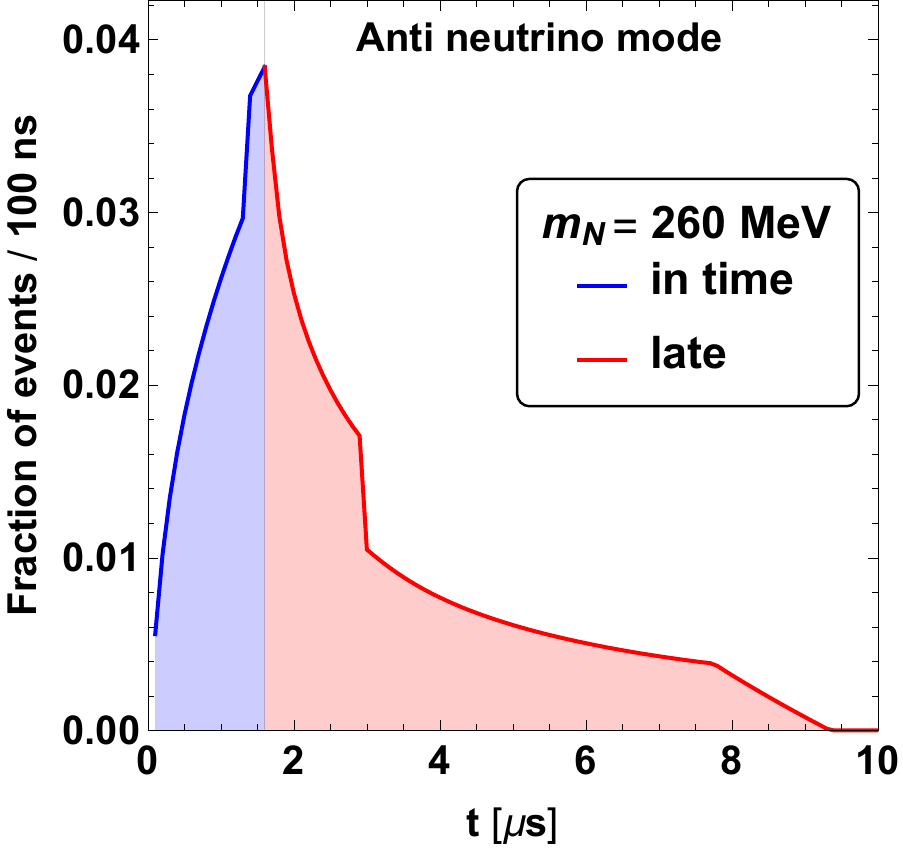}
    \caption{Time distribution of signal events for a sterile neutrino mass of 260~MeV in the neutrino (left) and antineutrino (right) beam mode. For the proton beam we assume a step-function of 1.6~$\mu$s duration. The zero of the time axis corresponds to the time when a neutrino produced at the onset of the beam traveling at the speed of light would arrive at the detector. The blue shaded region indicates the time window used for the analysis (1.6~$\mu$s); it contains 41\% (34\%) of all events in the neutrino (antineutrino) mode. }
    \label{fig:timespectrum}
\end{figure}

A heavy neutrino with momentum $p_N$ 
arrives at the detector at distance $L$ after a time
\begin{equation}
  t_N =  \frac{t_0}{\beta}\,, \qquad \text{with} \qquad
  t_0 = \frac{L}{c} \qquad \text{and} \qquad \beta = \frac{p_N}{E_N}\,,
    \label{eq:heavyNarrival}
\end{equation}
with the MiniBooNE baseline $L \simeq 540$ m.  The ultra-relativistic
light neutrinos all arrive after $t_0 \simeq 1.8 \,\mu$s, the heavy
neutrinos generally arrive later. In order to calculate the time
distribution of the events, we first convert the neutrino flux
$\Phi_N(p_N)$ into a function of time,
\begin{equation}
   \Phi(t) = \Phi_N(p_N) \left|\frac{dp_N}{dt}\right| 
   \label{eq:arrivaltime}
\end{equation}
with the Jacobian $|dp_N/dt| = p_N t/(t^2-t_0^2)$, which follows form eq.~\eqref{eq:heavyNarrival}.
In the decay model, an additional momentum dependence appears due to the effect of the Lorentz boost on the decay rate, which leads to a factor $m_N/p_N$, see eq.~\eqref{eq:Pdec} below. 
Finally, to construct the time spectrum we need to include the time structure of the proton beam, which we approximate with a step-function being non-zero from $t=0$ to $t=\delta t = 1.6\,\mu$s \cite{AguilarArevalo:2008qa}. Therefore, we obtain the time spectrum $T(t)$ in the following way:
\begin{equation}
    T(t) = \frac{1}{\delta t}\int_{t-\delta t}^t dt' \Phi(t') \frac{m_N}{p_N(t')}\,.
\end{equation}
We show the time distribution of the decay events inside the detector
for a typical heavy neutrino mass in fig.~\ref{fig:timespectrum}. The
contribution of the monochromatic peak from the stopped kaon decays is
visible in the discontinuous part of the red curve around $t=3\,\mu$s.
It is important to notice, that the neutrino appearance analysis from
MiniBooNE considers only events that occur between $t_0$ and
$t_0+1.6\,\mu$s after each beam spill \cite{BillLouis}. The fraction
of our heavy neutrino signal inside the analysis window is denoted in
blue in the figure, those that arrive after $t_0+ \delta t$ are too
late to be included and are denoted in red. The fraction of the events
inside the timing window is 41\% (34\%) in the neutrino
(antineutrino) mode.  Therefore, we predict a significant fraction of
delayed events. Those could be searched for in the MiniBooNE
data. Note that MiniBooNE records events within a time window of about
19.8~$\mu$s around each beam spill
(cf.\ ref.\ \cite{AguilarArevalo:2008qa}). A detailed investigation of
events in this time region can be a definite test of our model. Using
timing information to test heavy neutrino decays has been suggested
previously in ref.~\cite{Ballett:2016opr}.

\subsection{Event numbers, energy and angular spectra}
%
The number of heavy neutrinos that decay inside the detector are obtained by integrating over the heavy neutrino flux $\phi_N$, together with the probability $P_{\rm dec}$ that the long lived particles decay within its fiducial volume. Furthermore, a detection efficiency $\epsilon$ has to be included that is an empirical function of the signal energy, here approximated with the momentum of the photon, and the decays have to occur inside a timing window as discussed above.
These considerations are summed up in the following master formula:
\begin{align}
  N_{\rm decay} =& {\rm POT } \, \rho_\ell(m_N) \, {\rm Br}_{\nu\gamma} \, A_{\rm MB} \int dp_N \,\phi_N(p_N) \hat\epsilon(p_N) P_{\rm dec}(p_N) w_{\rm time}(p_N,m_N) \,. 
 \label{eq:master}
\end{align}
Here, POT denotes the number of protons on target, which is $12.84\,
(11.27)\times 10^{20}$ for the neutrino (antineutrino) mode. The
factor $\rho_\ell(m_N)$ has been defined in eq.~\eqref{eq:rho} and it
includes the mixing matrix element $|U_{\ell 4}|^2$ and the branching
ratio of the kaon decays into heavy neutrinos. Br$_{\nu\gamma} =
\GammaN / \Gamma_{\rm tot}$ is the branching ratio for the decay
$N\to\nu\gamma$, with $\Gamma_{\rm tot}$ being the total decay width
of $N$. In the relevant mass range we have $\Gamma_{\rm tot} \approx
\GammaN + \Gamma_\pi$ with $\Gamma_\pi$ given in
eq.~\eqref{eq:GammaSM}. Furthermore, $A_{\rm MB} = \pi (5\,{\rm m})^2$
is the effective area of the MiniBooNE detector, and
\begin{align}
  \hat\epsilon(p_N) = \int_{p_{\gamma,\rm min}}^{p_{\gamma,\rm max}}
  dp_\gamma \epsilon (p_\gamma) \frac{1}{\GammaN^{\rm lab}} \frac{d\GammaN^{\rm lab}}{dp_\gamma}    
\end{align}
is the MiniBooNE detection efficiency \cite{BillLouis} $\epsilon(p_\gamma)$ averaged over the photon momentum distribution for a given $p_N$. $P_{\rm dec}$ is the probability that the heavy neutrino decays inside the detector, and $w_{\rm time}$ is a timing-related weight.
Using the heavy neutrino arrival time $t_N$ from eq.~\eqref{eq:heavyNarrival} the latter is given by
\begin{equation}
  w_{\rm time}(p_N,m_N) = \left\{\begin{array}{ll}
  \frac{t_0 + \delta t - t_N}{\delta t} & \text{for } t_N<\delta t + t_0\\
  0 & \text{for } t_N\geq \delta t + t_0\,.
  \end{array}\right.
\end{equation}

For the decay probability we have
\begin{align}
  P_{\rm dec}(p_N) &= 
  e^{-L_1 \Gamma_{\rm tot} \frac{m_N}{p_N}} - 
  e^{-L_2 \Gamma_{\rm tot} \frac{m_N}{p_N}}  \label{eq:Pdec-exact}\\
  &\approx \Gamma_{\rm tot} \frac{m_N}{p_N} \Delta L \,, \label{eq:Pdec}
\end{align}
where $L_1, L_2$ denote the distance of the front and back ends of the detector from the beam production and we assume an effective value of $\Delta L \equiv L_2 - L_1 =8$ m. Here $\Gamma_{\rm tot}$ is the heavy neutrino width in the rest frame, a factor $m_N/E_N$ takes into account the boost into the lab frame of the detector, and $L_i \times E_N/p_N$ is the time the neutrino needs to reach the position $L_i$. In eq.~(\ref{eq:Pdec}) we have used an approximation, which holds for the MiniBooNE baseline of $L \approx 540$~m when $\Gamma_{\rm tot} \lesssim 10^{-15}$~MeV.
In this approximation the number of events is proportional to
$|U_{\ell 4}|^2  {\rm Br}_{\nu\gamma} \Gamma_{\rm tot} = |U_{\ell 4}|^2 \GammaN$.

We remark that in order to fit the observed shapes of the angular and
energy spectra, in principle we have to recast the energy of the
photon-induced Cherenkov shower (here assumed to be identical to the
momentum of the photon from the heavy neutrino decay) into the energy of
a light neutrino, $E_\nu$, assuming a quasi-elastic scattering
event. It turns out, however, that this recasting yields a relative difference on the percent level, such that we will keep the photon energy as our proxy for the reconstructed neutrino energy in the following, for simplicity.

Using the linear approximation for the decay probability \eqref{eq:Pdec}, and the differential decay widths from eqs.~\eqref{eq:dGammadp} and \eqref{eq:dGammadz}, we construct the predicted angular spectrum ${\cal A}(z)$ with $z\equiv \cos\theta$ and energy spectrum ${\cal E}(E_\nu)$ in the following way:
\begin{align}
{\cal A}(z) = & {\rm POT}\,\rho_\ell(m_N) \, A_{\rm MB} \Delta L \int d p_N \Phi_N(p_N) \frac{E_N}{p_N} w_{\rm time}(p_N,m_N)  \frac{d \Gamma^{\rm lab}_{N\to \nu \gamma}}{d z} \epsilon(p_\gamma(z))\,, \label{eq:angulardist} \\
{\cal E}(p_\gamma) = & {\rm POT}\, \rho_\ell(m_N) \, A_{\rm MB}\Delta L \int d p_N \Phi_N(p_N) \frac{E_N}{p_N} w_{\rm time}(p_N,m_N) \frac{d \Gamma^{\rm lab}_{N\to \nu \gamma}}{d p_\gamma} \epsilon(p_\gamma) \,,
\label{eq:energydist}
\end{align}
where the neutrino flux depends on the horn mode and includes forward and backward decays of the parent meson. 
The photon momentum is related to the scattering angle by
\begin{equation}
p_\gamma(z) = \frac{m_N^2}{2 (E_N - p_N \cos \theta)}\,.
\label{eq:pgamma}
\end{equation}

\begin{figure}
    \centering
    \includegraphics[width=0.6\textwidth]{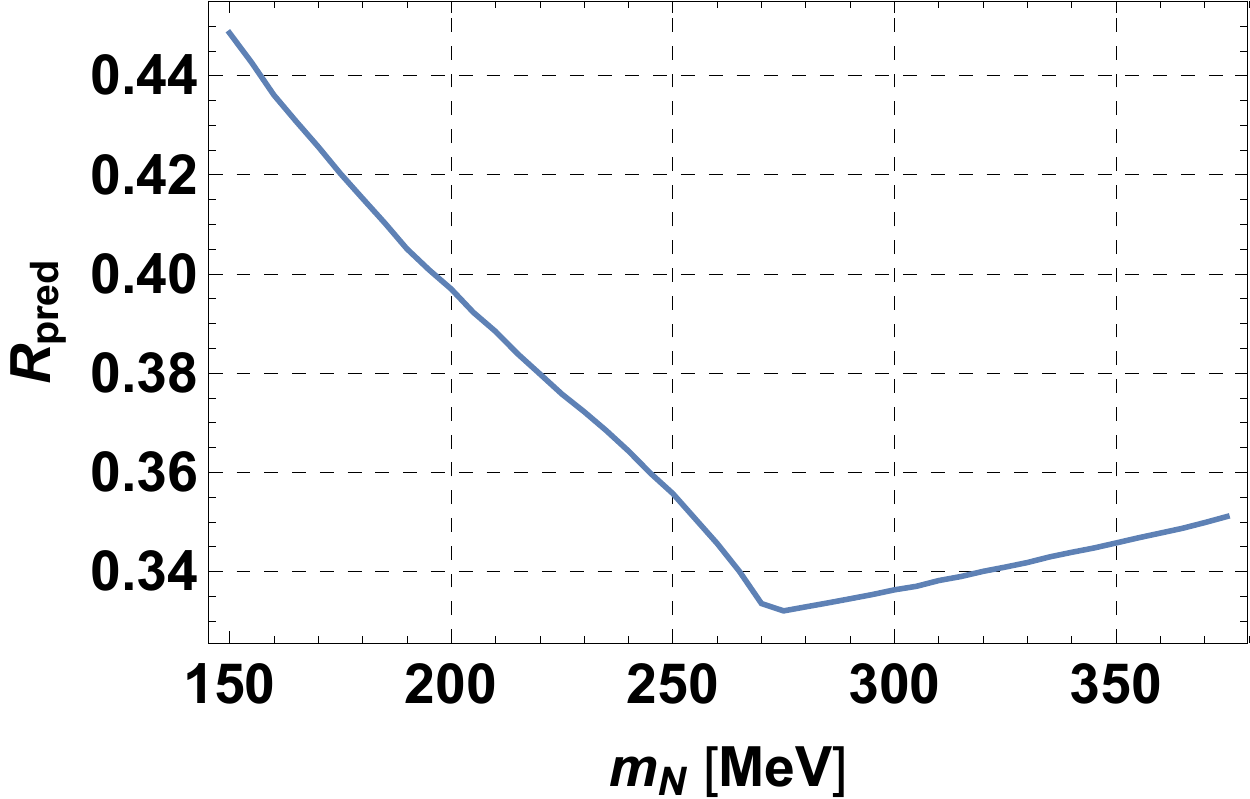}
    \caption{Predicted ratio of heavy neutrino events in the MiniBooNE detector for the horn being in antineutrino mode to the one in neutrino mode as a function of the heavy neutrino mass. 
    }
    \label{fig:ratio}
\end{figure}

In our model the ratio of signal events for the two horn polarisations ($R_{\rm pred}$) is determined from the corresponding fluxes, and can be calculated from eq.\ \eqref{eq:master} for a given heavy neutrino mass. We show $R_{\rm pred}$ as a function of $m_N$ in fig.\ \ref{fig:ratio} for the heavy neutrino being produced together with a muon.
The ratio is almost identical when the heavy neutrino is produced together with an electron, aside from the fact that larger values for $m_N$ are kinematically accessible.

\section{Fit to the data}\label{sec:fit}

In order to test our model we perform a fit to both, the angular and energy spectra. 
The data and the different background contributions are read from fig.\ 14 of ref.\ \cite{Aguilar-Arevalo:2018gpe}.
Ideally one would fit the angular and energy information simultaneously by using the 2-dimensional distribution of the data. Unfortunately this information is not available, and therefore we have to fit the energy and angular spectra separately and check if results are consistent.\footnote{Fitting the 1-dimensional spectra together would imply a double-counting of the same data.} 
In each case the fit is done fitting simultaneously both the neutrino and anti-neutrino spectra. 

Using the results of the previous section, we parameterize our model with two effective
parameters, which we chose to be $N_{\rm total}$ and the sterile
neutrino mass $m_N$. The predicted number of events in a given bin $i$ of the energy or angular data is given by $N_\nu f_i^{\nu}(m_N)$ and $N_{\bar\nu} f_i^{\bar\nu}(m_N)$ for the neutrino and anti-neutrino polarization, respectively. Here,
\begin{equation}
    N_\nu = \frac{N_{\rm total}}{1+R_{\rm pred}(m_N)} \,,\qquad
    N_{\bar\nu} = N_{\rm total}\frac{R_{\rm pred}(m_N)}{1+R_{\rm pred}(m_N)} \,,
\end{equation}
where $R_{\rm pred}(m_N)$ is the predicted ratio of events in the neutrino and anti-neutrino modes shown in fig.~\ref{fig:ratio}, and  $f_i^{\nu}(m_N)$, $f_i^{\bar\nu}(m_N)$ are the predicted relative contributions for each bin, normalized to 1. They are derived from the corresponding differential spectra given in eqs.~\eqref{eq:angulardist} and \eqref{eq:energydist}.

\begin{table}[t]
\centering
 \begin{tabular}{|c|c|c|}
  \hline
  Bkg contribution & $\nu$ mode & $\bar \nu $ mode \\ 
  \hline
  $\nu_e$ from $\mu$ & 0.24 & 0.3 \\
  $\nu_e$ from $K^{\pm}$ & 0.22 & 0.21 \\
  $\nu_e$ from $K^0$ & 0.38 & 0.35 \\
  $\pi^0$ miss & 0.13 & 0.10 \\
  $\Delta \to N \gamma$ & 0.14 & 0.16 \\
  dirt & 0.25 & 0.25\\
  other & 0.25& 0.25\\
  \hline
 \end{tabular}
 \caption{Relative uncertainties $\sigma_a$ and $\bar \sigma_a$ for the various background components, 
 taken from table~1 of ref.\ \cite{Aguilar-Arevalo:2018gpe}.
 Uncertainties for 'dirt' and 'other' are estimates. All background uncertainties are assumed to be uncorrelated. \label{tab:syst}}
\end{table}

\begin{figure}[t]
\centering
\includegraphics[width=0.67\linewidth]{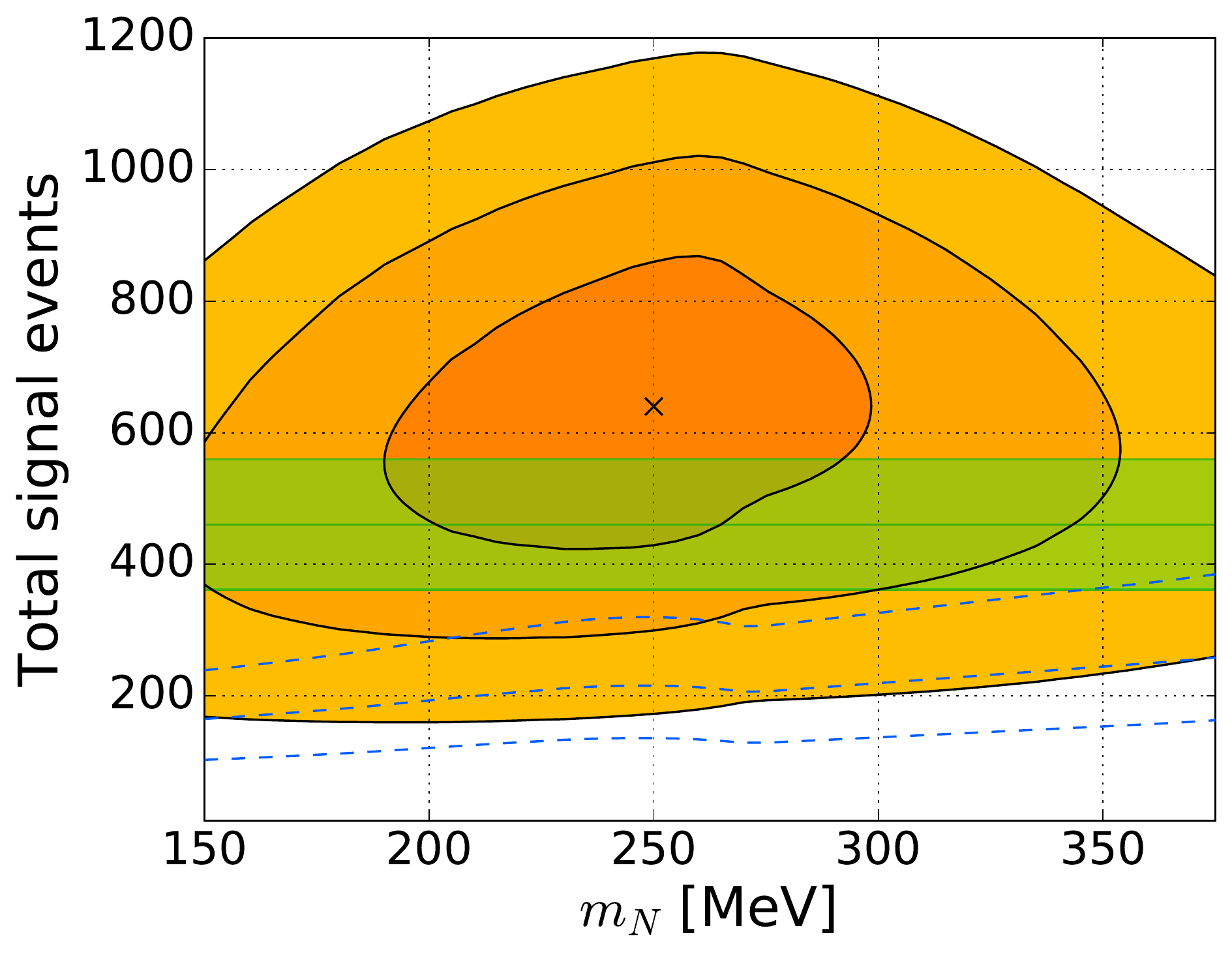}
\caption{Allowed regions at 1, 2, and 3$\sigma$ for the energy (orange
  regions) and angular (dashed blue curves) in the $N_{\rm total}$
  versus $m_N$ parameter space. The best fit of the energy spectral
  fit is indicated with a cross. The angular fit provides an upper
  limit on $N_{\rm total}$. In green we show the measured excess of
  events and its 1$\sigma$ uncertainty. We assume here heavy neutrino
  mixing with the muon; results for the electron are very similar.}
\label{2D chisq}
\end{figure}

We define the following $\chi^2$-function to perform the analysis:
\begin{eqnarray}
\chi^2(N_\nu,m_N) &=& \sum_i \frac{\left( O_i^\nu - b_a B_i^a - N_\nu f_i^\nu(m_N) \right)^2}{(\sigma_i^{\rm stat})^2 + (\sigma_i^{\rm syst})^2} \nonumber \\
&& + \sum_i \frac{\left( O^{\bar \nu_i} - \bar b_a \bar B^a_i - R_{\rm pred} N_\nu f_i^{\bar \nu}(m_N) \right)^2}{(\bar \sigma_i^{\rm stat})^2 + (\bar \sigma_i^{\rm syst})^2} \nonumber \\
&& + \sum_a \left( \frac{ b_a -1 }{\sigma_a} \right)^2 + \sum_a \left( \frac{ \bar b_a -1 }{\bar \sigma_a} \right)^2 \, .
\label{EQ: chisq min}
\end{eqnarray}
Here $i$ labels the angular or energy bins, $a$ labels the background contributions (sum over $a$ is implicit),
$O^\nu_i$ and $O^{\bar \nu}_i$ are the number of events in each bin $i$ for the $\nu$ and $\bar \nu$ mode respectively.
$B^a_i$ and $\bar B^a_i$ are the different $a$ background contributions in each bin $i$,
$b_a$ and $\bar b_a$ are the pull parameters that account for their uncertainty $\sigma_a$ and $\bar \sigma_a$,
which are taken from table~1 of ref.\ \cite{Aguilar-Arevalo:2018gpe}, see Tab.~\ref{tab:syst}.
Possible correlations are not taken into account.
Furthermore, a totally uncorrelated systematic uncertainty of 20\% is considered in each bin to account for possible spectral shape uncertainties:
$\sigma^{\rm syst}_i (N,m_N) = 0.2 N_\nu f^{\nu}_i(m_N)$ and
$\bar\sigma^{\rm syst}_i (N,m_N) = 0.2 N_{\bar\nu} f^{\bar \nu}_i(m_N)$.

The results of the angular and energy analyses are shown in
figure~\ref{2D chisq}. To be specific, we assume the production mode
$K\to N\mu$, results for $K\to Ne$ are very similar.  The energy
spectrum provides a best fit point at $m_N = 250$~MeV and $N_{\rm
  total} = 640$ with closed allowed regions.  At 68\% confidence level
our fit allows the masses to vary between 190 MeV and 295 MeV and
normalisations between 425 and 865 events, consistent within 1$\sigma$
with the observed number of excess events $N_{\rm obs} = 460.5\pm
95.8$, as indicated by the green band in the plot. Note that this
comparison is only indicative, since the data used in our fit
(taken from fig.~14 of ref.~\cite{Aguilar-Arevalo:2018gpe}) uses a
lower energy threshold and therefore the number of excess events is
somewhat larger, consistent with our best fit value. The best fit
point has $\chi^2_{\rm min}/{\rm dof} = 58.1/36$ which corresponds to
a $p$ value of about 1\% (see discussion below). In contrast, the
angular spectrum only provides an upper bound on $N_{\rm total}$ which
is in some tension with the energy fit.

\begin{figure}
\centering
\includegraphics[width=0.49\linewidth]{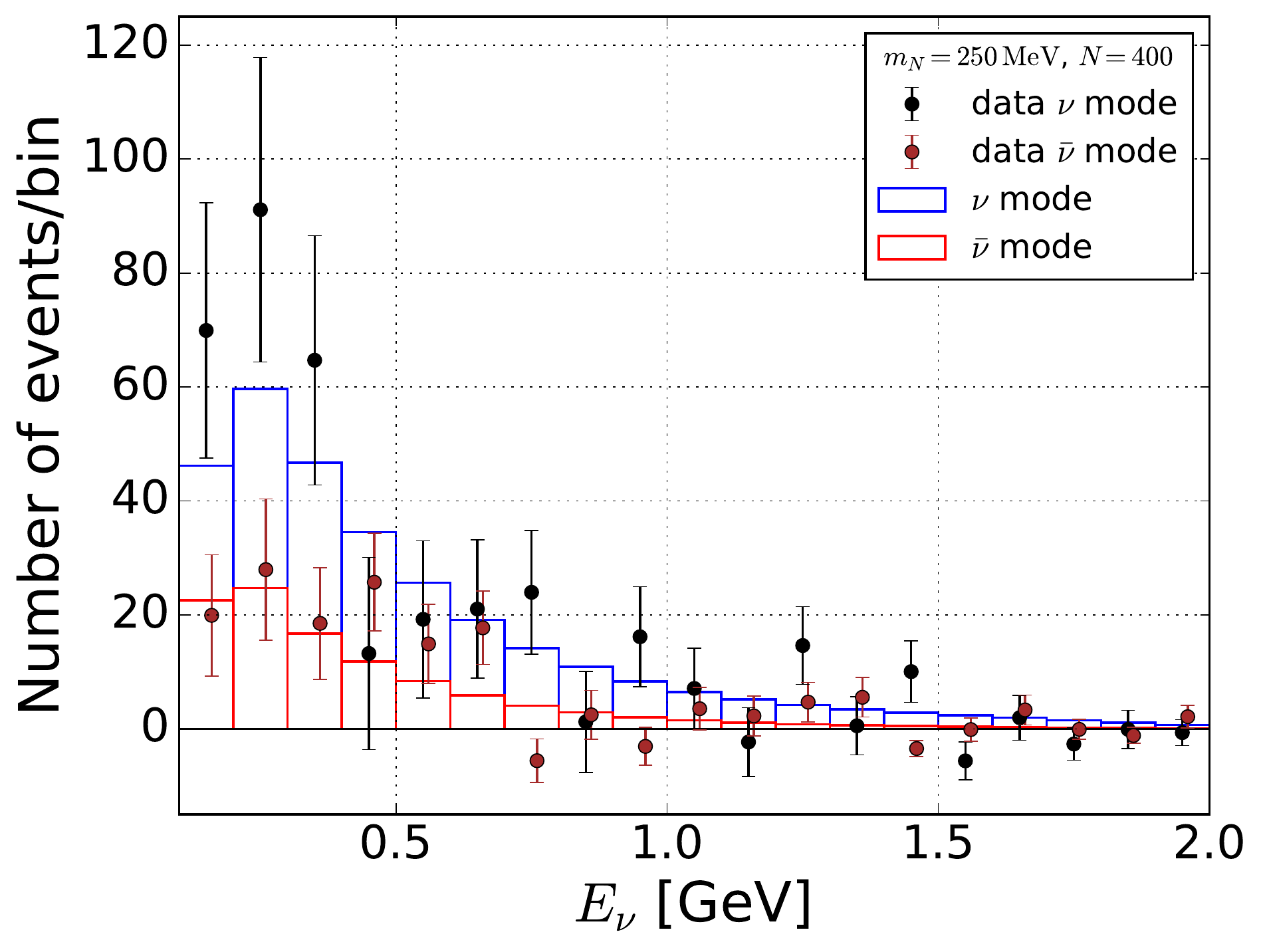}
\hfill
\includegraphics[width=0.49\linewidth]{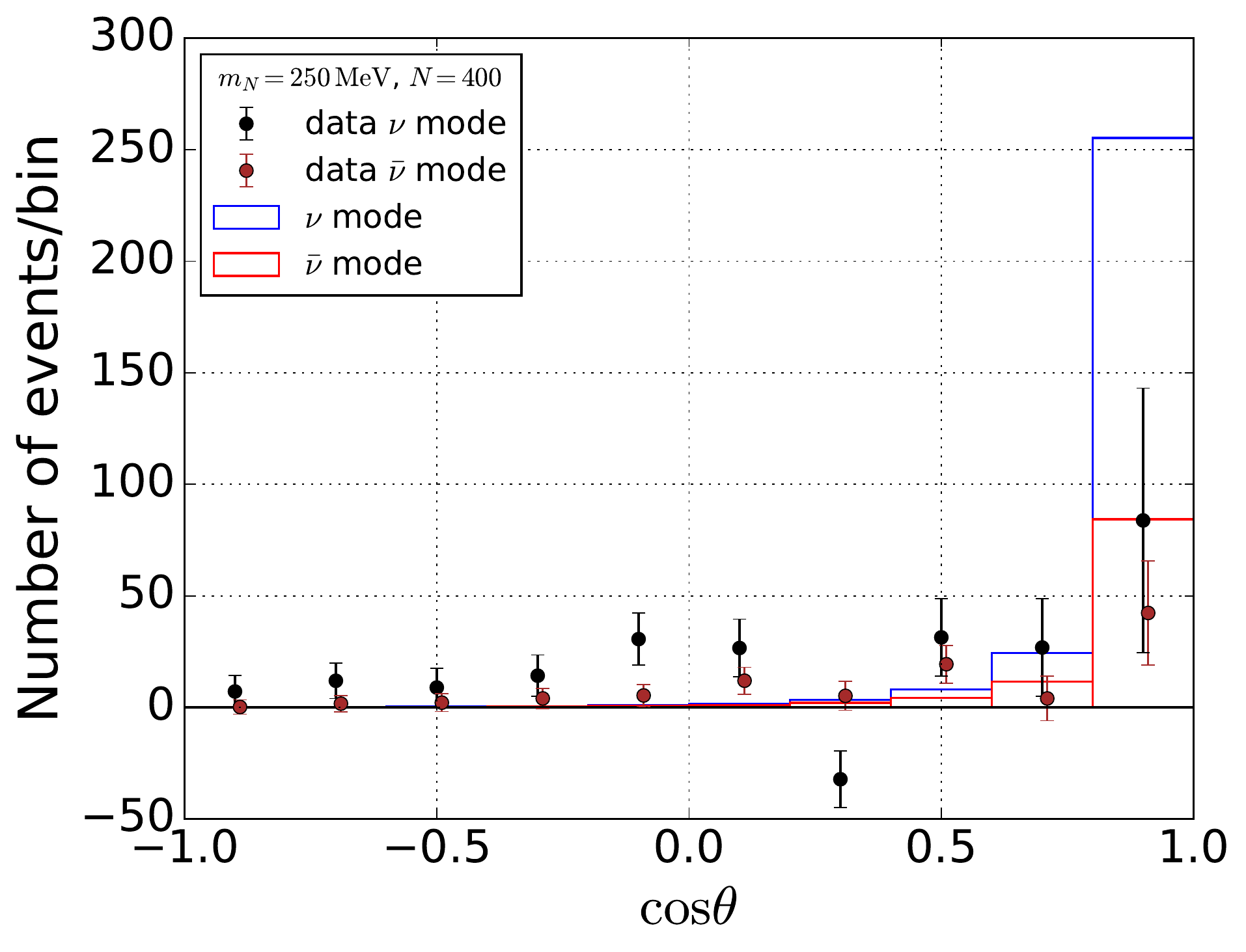}
\caption{Predicted energy (left) and angular (right) spectra
  for $m_N = 250 \, {\rm MeV}$ and $N_{\rm total} = 400$.
The dots are the data with background subtracted, where each background component is re-scaled by the corresponding pull parameter obtained from the fit. Error bars include statistical and systematical errors.}
\label{energy and angular spectra}
\end{figure} 

In order to investigate the quality of the fit we show in
figure~\ref{energy and angular spectra} the predicted energy and
angular spectra for $m_N=250$~MeV and $N_{\rm total}$ fixed to 400,
chosen within the 1$\sigma$ range of the observed value.  From the
left panel we see that our model explains well the excess events in
the energy spectrum. A significant contribution to the $\chi^2$ comes
from bins above 1~GeV, where a signal is neither observed nor
predicted. This explains the rather low $p$-value of only 1\%. The
right panel shows that the angular shape for the anti-neutrino mode is
in good agreement with the observations, while the signal is somewhat
too much forward peaked in the neutrino mode. From comparing the
neutrino-mode spectra in the two panels (blue histograms), the tension
between energy and angular fit is apparent. While the energy spectrum
would prefer to increase the normalization, this would clearly worsen
the prediction in the forward angular bin.  Note however, that the
largest contribution to the angular $\chi^2$ comes from the three bins around
$\cos\theta=0$, including the one with the downward fluctuation. Those
bins are difficult to explain by any smooth function.

A general discussion of the angular event distribution in decay models
can be found in ref.~\cite{Jordan:2018qiy}. We stress that to
definitely assess the viability of our model a joint energy and
angular fit should be performed, including detailed acceptances and
efficiencies suitable to our signature. Let us also mention that both
the timing cut and the implementation of the angular acceptance is
important to predict the angular shape, since both affect mostly the
signal from the decay of ``slow'' neutrinos, which give the main
contribution to events with $\cos\theta < 1$. In
appendix~\ref{app:timing} we show the fit results without imposing the
1.6$\mu$s timing cut, which leads to an improved angular fit. Below
we proceed under the assumption that our model does provide an
acceptable fit to MiniBooNE data.


\section{Discussion of results}\label{sec:results}

\subsection{Available parameter space of the model}

\begin{figure}[t]
    \centering
    \includegraphics[width=0.47\textwidth]{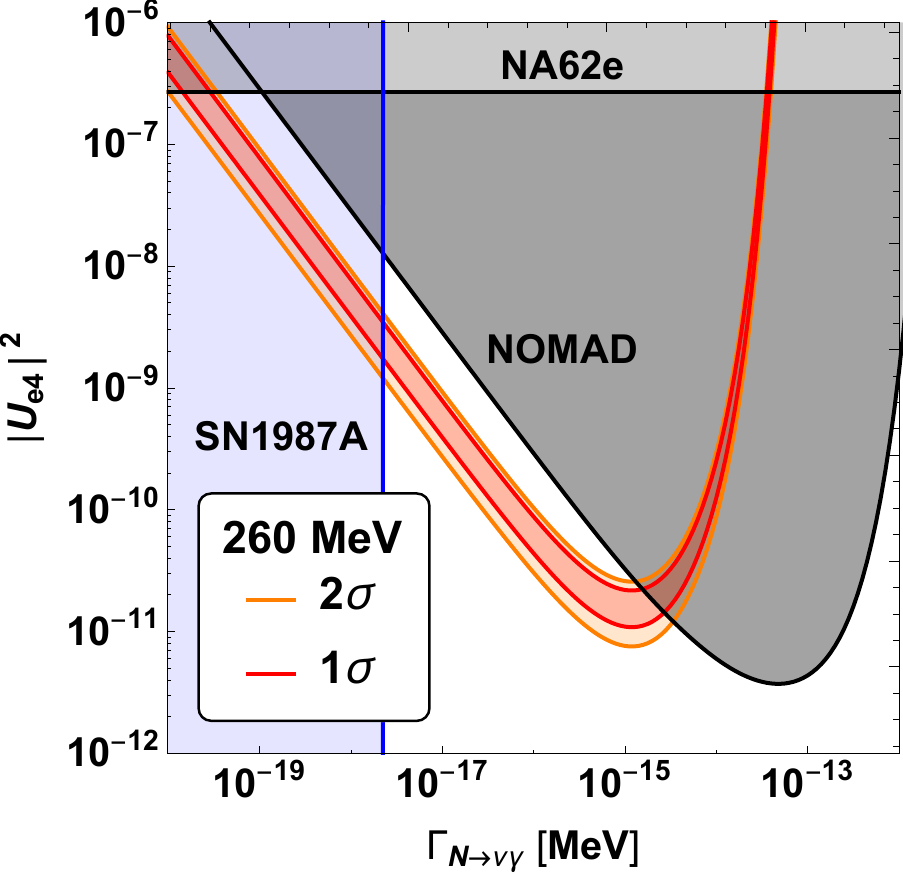}\quad
    \includegraphics[width=0.47\textwidth]{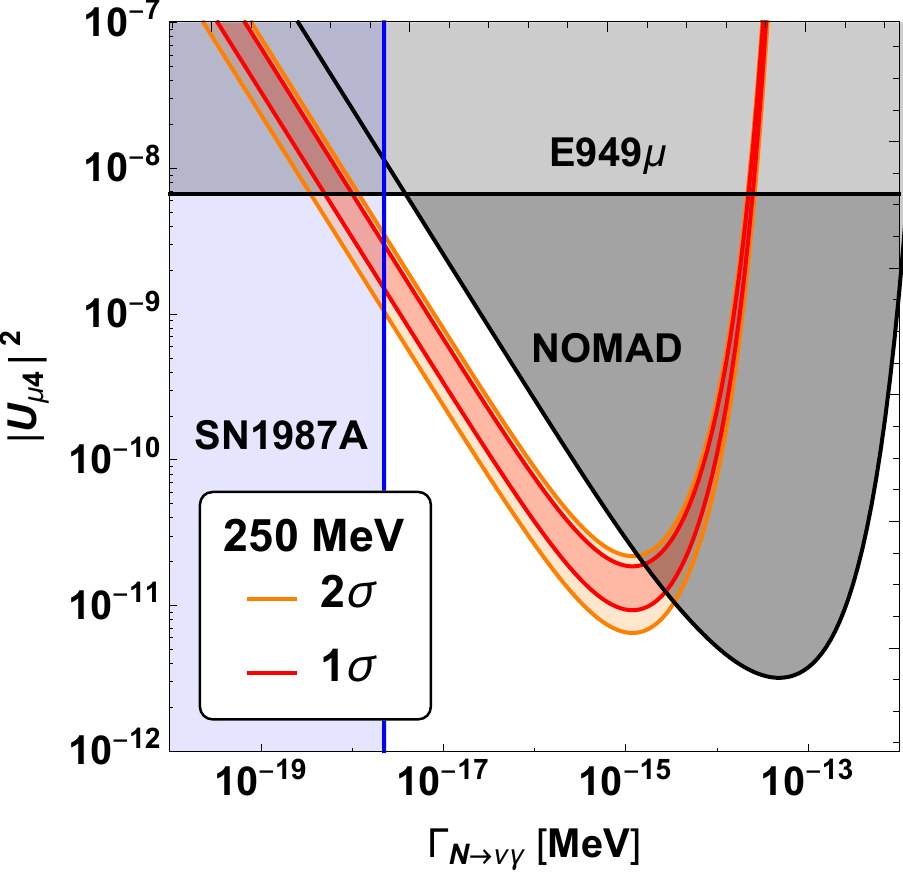}
    \caption{Parameter region in the plane of $\GammaN$ and $|U_{\ell
        4}|^2$ that is consistent with the observed MiniBooNE excess
      at 1 and 2$\sigma$ for the neutrino mass fixed at the best fit
      point. The left (right) plot assumes that $N$ is produced from a
      kaon decay with an associated electron (muon) and corresponds to
      $m_N = 260 \, (250)$~MeV.  Also shown are upper limits on
      $|U_{\ell 4}|^2$ from NA62 \cite{CortinaGil:2017mqf} and E949
      \cite{Artamonov:2014urb} and the region excluded by NOMAD from
      the search in ref.~\cite{Kullenberg:2011rd}, interpreted in our
      model, as well as the region for $\GammaN$ disfavoured by
      SN1987A~\cite{Magill:2018jla}.}
    \label{fig:parameterspace-U-Gamma}
\end{figure}

\begin{figure}
\centering
\includegraphics[height=0.45\textwidth]{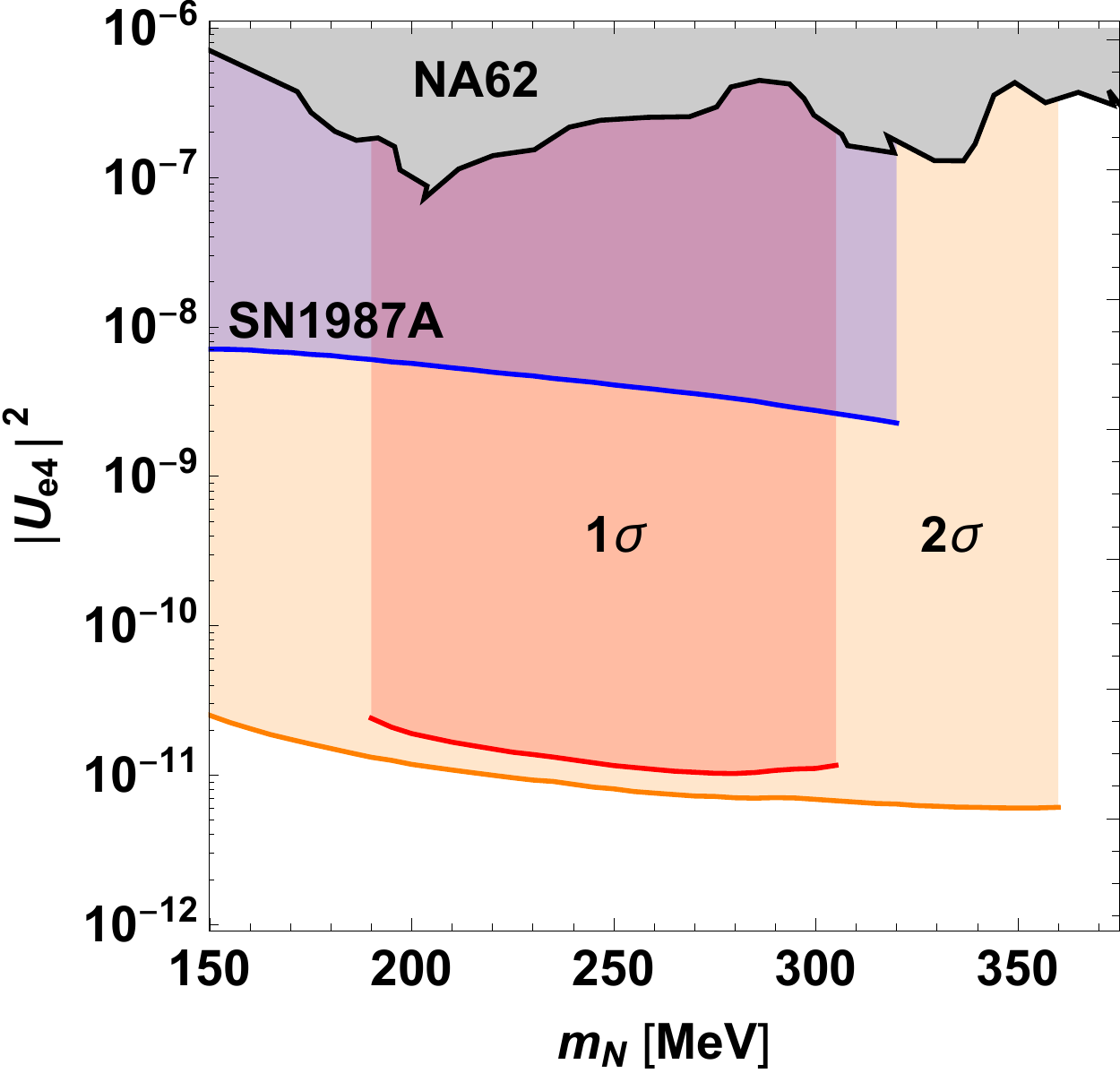}\qquad
\includegraphics[height=0.45\textwidth]{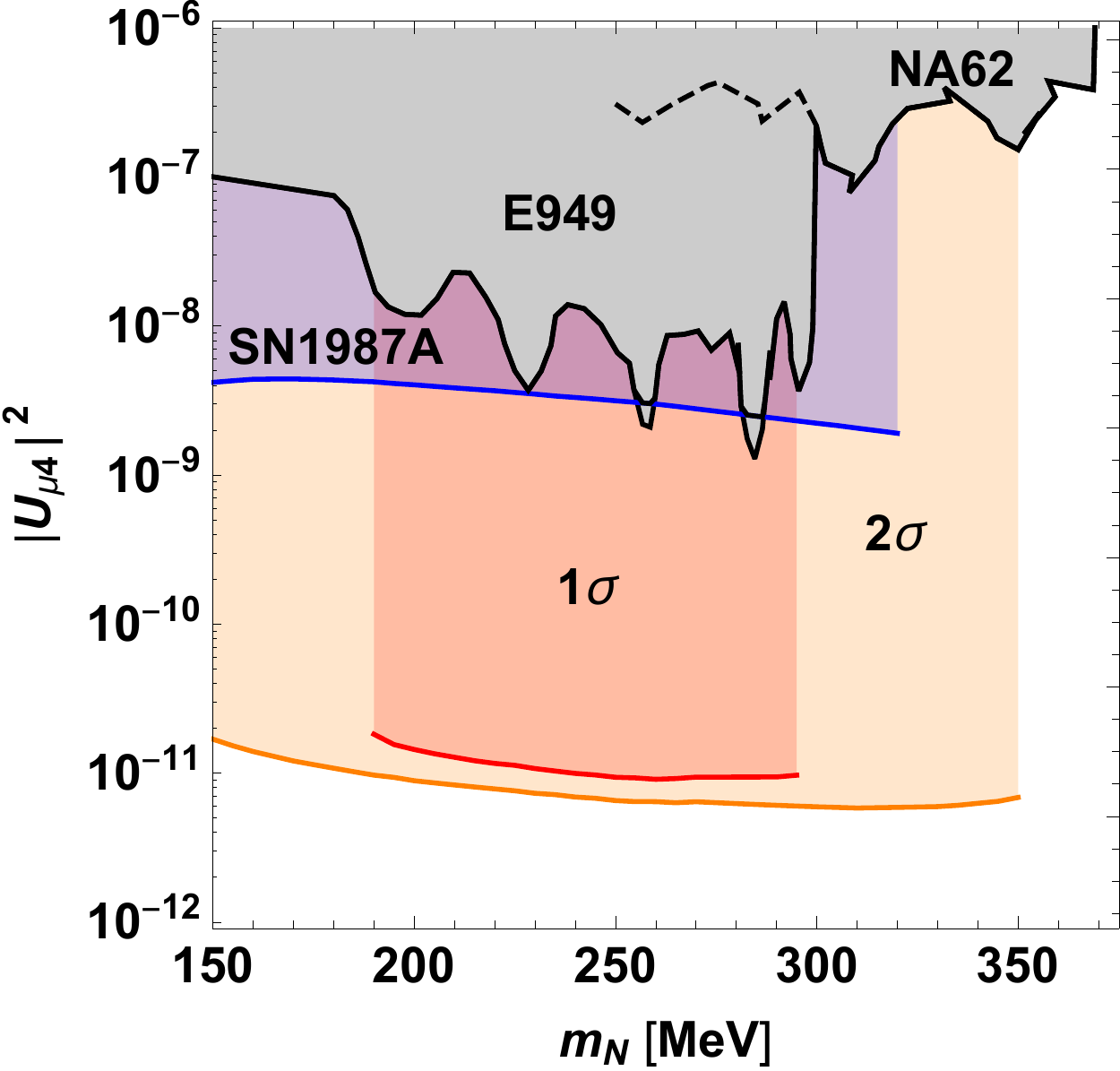}
\caption{Parameter region in the plane of $m_N$ and $|U_{\ell 4}|^2$
  that is consistent with the observed MiniBooNE excess at 1 and
  2$\sigma$. For each point in the allowed region the decay rate
  $\GammaN$ can be chosen such that the observed MiniBooNE events can
  be obtained. For the left (right) panel the heavy neutrino is
  produced by $K\to e N \, (\mu N)$. Excluded parameter space from
  peak searches in the kaon decay spectra of electron and muon from
  the NA62 \cite{CortinaGil:2017mqf} and E949 experiments
  \cite{Artamonov:2014urb} is shown as gray shaded regions.  The
  regions disfavoured by SN1987A constraints on
  $\GammaN$~\cite{Magill:2018jla} are shown as blue shaded regions.}
\label{fig:parameterspace}
\end{figure}

By using eq.~\eqref{eq:master}, the total number of events determined
by the fit above can be translated into the parameter space given by
the neutrino mixing parameter $|U_{\ell 4}|^2$ and the heavy neutrino
decay rate $\GammaN$. In fig.~\ref{fig:parameterspace-U-Gamma} we show
the 1 and $2\sigma$ contours for those two parameters for the neutrino
mass fixed at the best fit point. The straight part on the left side
corresponds to the linear approximation for the decay probability,
eq.~\eqref{eq:Pdec}, where event numbers are proportional to the
product $|U_{\ell 4}|^2 \GammaN$. The linear approximation breaks
down when the decay length becomes shorter than the MiniBooNE baseline
and most of the neutrinos decay before reaching the detector. This
leads to the upturn of the allowed region visible in the plots for
decay rates $\GammaN \gtrsim 10^{-15}$~MeV. This value depends only
weakly on $m_N$ and defines a minimum value of $|U_{\ell 4}|^2$ needed
to explain the excess of roughly $2\times 10^{-11}$.  The lower limit
on $|U_{\ell 4}|^2$ is shown as a function of the heavy neutrino mass
in fig.~\ref{fig:parameterspace}. The dark and light orange shaded
regions correspond to the 1$\sigma$ and 2$\sigma$ range for $m_N$ as
shown in fig.~\ref{2D chisq}. Note that we do not consider masses
below 150~MeV in order to avoid $N$ production due to pion decays. We
see that the excess can be explained by a wide range of values for the
mixing and for the decay rate. Let us now consider other constraints
on those parameters.

\paragraph{NOMAD:} The single photon signature predicted in our model can be searched for
in various other neutrino experiments. A rather sensitive search comes
from the NOMAD experiment at CERN. An overview of the experiment is
given in ref.~\cite{Vannucci:2014wna}.  An analysis searching for
single photon events (motivated by the MiniBooNE
observation) yields 78 observed events in forward direction versus
$76.6 \pm 4.9 \pm 1.9$ expected, which was interpreted as a null
result and an upper bound of 18 events at 90\%~CL has been set on
single photon events~\cite{Kullenberg:2011rd}. We can interpret this
bound as a limit within our model.
We use the kaon-produced muon neutrino flux from
ref.\ \cite{Astier:2003rj} and construct the heavy neutrino flux
as described in appendix~\ref{sec:flux}. The number of
heavy neutrino decays in the detector is estimated as in
eq.~\eqref{eq:master}.  The POT is $2.2\times 10^{19}$ and we use a
constant reconstruction efficiency of 90\%, an analysis efficiency of
8\%, and a trigger efficiency of 30\%~\cite{Kullenberg:2011rd}. In
order to take into account an analysis cut on the observed energy we
consider only heavy neutrinos with momentum greater than 1.5~GeV.
We do not apply any time window for the events.

The parameter space excluded by NOMAD by requiring that less than 18
events are predicted is shown as the dark gray shaded region in
fig.~\ref{fig:parameterspace-U-Gamma}. Since the baseline of NOMAD is
shorter than MiniBooNE and the neutrino energies are higher, the decay
rate for which neutrinos start decaying before reaching the detector
is shifted to higher values of $\GammaN$ for NOMAD compared to MiniBooNE and NOMAD
excludes the ``non-linear'' part of the parameter space. In the linear
regime for the decay probability the NOMAD bound is always consistent
with the value $|U_{\ell 4}|^2 \GammaN$ required to explain
MiniBooNE. We have checked that the predicted number of events in
NOMAD is about $5.4\times 10^{-3}$ times the signal events in
MiniBooNE, with very little dependence of this number on $m_N$ within
the interesting range. Therefore, the NOMAD bound limits the available
parameter space to the linear regime but does not provide a further
constraint on the range of the parameters.

\paragraph{Limits from kaon experiments.}
Due to the long lifetimes of the heavy neutrinos the vast majority of
the produced $N$ decay outside the detectors in most kaon
experiments. However, an observable feature of their presence is given
by an additional peak in the spectrum of the lepton from the decaying
kaon. The NA62 experiment has recently published a search for heavy
neutral leptons that are produced in kaon decays.  Not observing
candidates for kaon decays into heavy neutrinos, they placed upper
limits at the 90\% CL of around $|U_{\ell 4}|^2 \sim 10^{-7}$ for
$\ell=e,\mu$ and heavy neutral leptons with masses between 170 and 448
MeV for $\ell=e$ and between 250 and~373 MeV for $\ell=\mu$
\cite{CortinaGil:2017mqf}.  Earlier searches for heavy neutrinos from
the E949 experiment studied the muon spectra from about $10^{12}$
stopped kaon decays.  In their analysis, the collaboration derived the
still most stringent upper limits at the 90\%~CL on $|U_{\mu 4}|^2$
down to $10^{-9}$ for heavy neutrinos with masses between 175 and 300
MeV \cite{Artamonov:2014urb}.  We show the region excluded by E949 and
NA62 in figs.~\ref{fig:parameterspace-U-Gamma} and
\ref{fig:parameterspace} as a gray area.\footnote{Recently NA62 has
  presented preliminary updated limits \cite{NA62-talk}. They are in
  the range $|U_{\mu 4}|^2 < 2\times 10^{-8}$ for $220\,{\rm MeV}
  \lesssim m_N \lesssim 370$~MeV, and $|U_{e 4}|^2 < 2\times 10^{-9}$
  for $150\,{\rm MeV} \lesssim m_N \lesssim 400$~MeV.}

\bigskip

To summarize sofar, as visible in fig.~\ref{fig:parameterspace}, several
orders of magnitude in mixing are available to explain the MiniBooNE
excess in this model. For a fixed value of $m_N$, for each value of
$|U_{\ell 4}|^2$ in the allowed range, the value of the decay rate can
be adjusted such that the event numbers are kept
constant. Requiring $N_{\rm total} = 400$ events in MiniBooNE we have approximately
\begin{equation}\label{eq:Gamma-fit}
  \GammaN \simeq 3\times 10^{-17} \, {\rm MeV} 
    \left(\frac{10^{-10}}{|U_{\ell 4}|^2} \right)
  \left(\frac{250 \, {\rm MeV}}{m_N} \right)^{2.3}
  \left(\frac{N_{\rm total}}{400} \right) \,,
\end{equation}
where we have used the linear approximation for the decay probability and
the fact that then event numbers are proportional to the product
$|U_{\ell 4}|^2 \GammaN$.
The power of the mass dependence has been obtained by fitting
the numerical result with a power law and it is rather accurate in the
range $150 \, {\rm MeV} < m_N < 300\,{\rm MeV}$.

\paragraph{Constraint from SN1987A.} As discussed in ref.~\cite{Magill:2018jla}, a heavy neutrino interacting via the operator \eqref{eq:operator2} may contribute to the cooling rate of a supernova. In order to be consistent with the neutrino observation of SN1987A, too fast cooling has to be avoided. This argument can be used to disfavour certain regions in the parameter space of $m_N$ and $\GammaN$. In the parameter region of our interest those considerations lead to a lower bound on the decay rate of approximately~\cite{Magill:2018jla}
\begin{equation}\label{eq:SN}
  \GammaN > 2.4\times 10^{-18} \, {\rm MeV} \left(\frac{250\, {\rm MeV}}{m_N}\right) \,,
  \qquad (50\,{\rm MeV} \lesssim m_N \lesssim 320\,{\rm MeV}) \,.
\end{equation}
For decay rates fulfilling this bound, the heavy neutrino is sufficiently trapped inside the supernova to avoiding too fast cooling. The bound shown in eq.~\eqref{eq:SN} holds in the relevant mass range for our scenario, up to $m_N \approx 320$~MeV; heavier neutrinos are gravitationally trapped inside the supernova \cite{Dreiner:2003wh}. The region disfavoured by the bound \eqref{eq:SN} is shown in figs.~\ref{fig:parameterspace-U-Gamma} and \ref{fig:parameterspace} as blue shaded region, where in order to translate the bound into $|U_{\ell 4}^2|$ as shown in fig.~\ref{fig:parameterspace} we assume our explanation of the MiniBooNE events, using the relation \eqref{eq:Gamma-fit}. We see that in the mass range where the SN bound applies, the mixing is limited to $10^{-11} \lesssim |U_{\ell 4}^2| \lesssim {\rm few}\times 10^{-9}$, while for $m_N > 320$~MeV values of $|U_{\ell 4}^2|$ up to the kaon bounds of order $10^{-7}$ are allowed. Once these constraints from the magnetic moment operator are imposed, the supernova limits on mixing from ref.~\cite{Dolgov:2000jw} are satisfied; they disfavour the region $m_N \lesssim 100$~MeV and $|U_{\ell 4}^2| \gtrsim 10^{-8}$.

\bigskip

By comparing the decay rate from eq.~\eqref{eq:Gamma-fit} with the
mixing induced decay rate in pions given in eq.~\eqref{eq:GammaSM} we
find that $\Gamma_\pi \ll \GammaN$ for
\begin{align}\label{eq:cond}
  |U_{\ell 4}|^2  \ll 10^{-7}
 \left(\frac{250 \, {\rm MeV}}{m_N} \right)^{2.65}
  \left(\frac{N_{\rm total}}{400} \right)^{1/2} \,. 
\end{align}
We see from fig.~\ref{fig:parameterspace} that for the largest allowed
mixing angles in the high-mass region this condition may be violated.
In this case, the decays $N\to \nu\pi^0$ and $N\to\ell^\pm \pi^\mp$ can provide an additional observable signature.
Note, however, that in the region where the linear approximation
breaks down, $\Gamma_\pi \ll \GammaN$ is satisfied and we can use
$\Gamma_{\rm tot} \approx \GammaN$ for calculating the decay
probability according to eq.~\eqref{eq:Pdec-exact}.

\subsection{Other searches and tests of the model}

\paragraph{The PS191 and E816 experiments:}
The dedicated PS191 experiment searched for displaced vertices from
the decay of heavy neutrinos in the mass range from a few tens of MeV
to a few GeV.  Not having found such vertices PS191 placed limits on
the mass-mixing parameter space
\cite{Vannucci:1985vs,Bernardi:1987ek}.  It is important to notice
that these limits are not applicable in the here considered model,
because the decays of the heavy neutrino into a photon and a light
neutrino do not produce a visible vertex in the decay volume.

We remark that the experiment observed an excess of electron-like
events \cite{Bernardi:1986hs}, which was interpreted as
electron-neutrino scatterings in the calorimeter, but might be also
induced by the photons from the $N$ decay in our model.  This finding
is backed up by the PS191 successor at BNL, the experiment E816
\cite{Astier:1989vc}.  Unfortunately the collaborations do not provide
the details on the kaon flux, such that we cannot quantify the
respective signal strengths in our model.

\paragraph{LSND and KARMEN:}
The LSND \cite{Aguilar:2001ty} and KARMEN \cite{Armbruster:2002mp} experiments produce neutrinos from muon decay at rest and therefore heavy neutrinos with masses of $\gtrsim 100$~MeV will not be produced. The interactions of the 800~MeV proton beam with the target might produce a few slow-moving kaons, which could give rise to a heavy neutrino flux that is small compared to the one at MiniBooNE. Furthermore, the standard search in LSND and KARMEN requires a coincidence signal between a prompt positron and delayed neutron capture from the $\bar\nu_e$ inverse beta decay process, which is rather distinctive from the pure electro-magnetic signal induced by the single photon decay in our model. Therefore, we predict a negligible event rate in those experiments.

\paragraph{T2K, NO$\nu$A, and other running neutrino experiments:}
Modern neutrino detectors, such as the near detectors of NO$\nu$A, T2K
are generally not expected to confuse a single photon with charged
current electron neutrino scattering due to their more sophisticated
detectors. Recently the T2K collaboration published results for a
search for heavy neutrinos \cite{Abe:2019kgx} by looking for events
with two tracks, for instance from the decays $N\to \mu^\pm \pi^\mp$
or $N\to \ell^\pm \ell^\mp$.  The limits, comparable to those from
PS191 and E949, are not applicable to our model. A search for single
photon events in T2K has been published recently in
\cite{Abe:2019cer}. We have roughly estimated the sensitivity of this
result to our model and found that the resulting limits are weaker than the ones
from NOMAD discussed above.

It is important to realize that the signal of our model mimics
neutral current produced $\pi^0$ decays where one photon was not
reconstructed, which may interfere with the control regions of any analysis
and affect results in a non trivial way \cite{Arguelles:2018mtc}.  An
analysis that searches for single photons in the data in all running
neutrino experiments may be able to shed light on the MiniBooNE
excess. The relative signal strength between experiments is fixed by
the fluxes and allows to reject the hypothesis.

\paragraph{ISTRA+:}
The ISTRA+ experiment searched for and excluded the process $K^\pm\to
\mu^\pm N$, $N\to \nu \gamma$ for 30 MeV $\leq m_N \leq$ 80 MeV
\cite{Duk:2011yv} and for very short neutrino lifetimes.  With about
300 million events on tape, the experiment could in principle be
sensitive to heavy neutrinos in the here considered mass range.

\paragraph{The Fermilab short-baseline neutrino program:} 

The short-baseline neutrino (SBN) program at Fermilab consists of
three liquid argon detectors in the booster neutrino beam line: SBND,
MicroBooNE, and Icarus \cite{Antonello:2015lea, Machado:2019oxb}, with
the MicroBooNE detector already running and producing results. A
sensitivity study to heavy neutrino decays, including the photon decay
mode has been performed in \cite{Ballett:2016opr}, see also
\cite{Alvarez-Ruso:2017hdm}. Liquid argon detectors will be very
suitable to search for the signal predicted here, since such detectors
can discriminate photons from electrons. The main characteristics of
the three detectors are summarized in tab.~\ref{tab:sbn}. Since they
are located in the same beam as MiniBooNE we can roughly estimate the
expected number of events by scaling with the proportionality factor
\begin{equation}\label{eq:scaling}
  {\rm POT} \times V / L^2 \,,
\end{equation}
where $V$ is the detector volume and $L$ the distance of the detector
from the neutrino source. Note that the simple scaling with this
assumes that the linear approximation for the decay probability holds
for all baselines.  In the table we give this scaling factor for each
experiment relative to MiniBooNE (``Ratio''). Assuming 400 signal
events in MiniBooNE, we can predict then the expected number of events
by multiplying with this ratio. As is clear from the last row in
tab.~\ref{tab:sbn} a significant number of events is predicted for
each of the three detectors, under the quoted assumptions on the
available POT \cite{Ballett:2016opr}.

\begin{table}
  \centering
  \begin{tabular}{|l|c|ccc|}
    \hline
    & MiniBooNE & SBND & MicroBooNE & Icarus \\
    \hline
    POT / $10^{20}$  & 24 & 6.6 & 13.2 & 6.6 \\
    Volume / m$^3$  & 520 & 80 & 62 & 340 \\
    Baseline / m    & 540 & 110 & 470 & 600 \\
    Ratio           &     & 1 & 0.09 & 0.15 \\
    Events          & 400 & 400 & 35 & 58 \\
    \hline
  \end{tabular}
  \caption{Benchmark characteristics of the three SBN detectors
    \cite{Antonello:2015lea, Machado:2019oxb} compared to
    MiniBooNE. We assume the same POT as quoted in
    \cite{Ballett:2016opr}. For MiniBooNE we sum the POT in neutrino
    and antineutrino modes. The row ``Ratio'' indicates the ratio of
    signal events relative to MiniBooNE based on the scaling with the
    factor in eq.~\eqref{eq:scaling}. In the row ``Events'' we give the
    predicted number of events assuming 400 signal events in
    MiniBooNE. \label{tab:sbn}}
\end{table}

\paragraph{Atmospheric and solar neutrinos.} The magnetic moment operator \eqref{eq:operator2} can lead to the up-scattering of atmospheric or solar neutrinos to the heavy neutrino, which can give observable signals in IceCube \cite{Coloma:2017ppo} or dark matter detectors \cite{Shoemaker:2018vii}, respectively. The latter can test heavy neutrinos with mass below $\sim 10$~MeV. The sensitivities of IceCube derived in ref.~\cite{Coloma:2017ppo} from atmsopheric neutrinos are in the relevant mass range, but are about one order of magnitude too weak in $\GammaN$ to start constraining the parameter space relevant for our MiniBooNE explanation.

\section{Conclusions}\label{sec:conclusions}

We presented a model with a heavy neutrino of mass around 250~MeV that
is produced from kaon decays at the beam-target interaction via the
mixing $|U_{\ell 4}|^2$ ($\ell=e,\mu$) and decays after traveling over
several hundred meters into a light neutrino and a single photon via
an effective interaction.  We demonstrated that it is possible to
account for the event numbers and spectral shape of the electron-like
excess in the MiniBooNE data under the assumption that single photons
are indistinguishable from single electrons. Some tension appears for
the angular distribution of excess events, which are somewhat too much
forward peaked. A quantitative assessment of this tension requires a
dedicated analysis of MiniBooNE data including a careful treatment of
angular and timing acceptances.

The excess events can be explained for a wide range of mixing
parameters of roughly $10^{-11} \lesssim |U_{\ell 4}|^2 \lesssim
10^{-7}$, consistent with existing bounds, see
fig.~\ref{fig:parameterspace}. The model makes clear predictions and
can be tested in the following way:

\begin{itemize}
\item {\bf Delayed events in MiniBooNE: }Due to the non-negligible
  mass of the heavy neutrino, we predict a characteristic time
  structure of the signal with a significant fraction (up to 60\%) of
  events outside the time window corresponding to the time structure
  of the beam and assuming speed of light for the propagation to the
  detector. Therefore, the model can be tested by looking for delayed
  events in MiniBooNE data.

\item {\bf Single photon events in SBN detectors:}  We predict a
  sizable number of single photon events in all three liquid argon
  detectors of the Fermilab short-baseline neutrino program (SBND,
  MicroBooNE, ICARUS). These detectors have good photon
  identification abilities and should be able to confirm or refute our
  hypothesis.
\end{itemize}

Assuming that the decay $N\to\nu\gamma$ is induced by the dimension-5
operator of the magnetic moment type, see eqs.~(\ref{eq:operator2},
\ref{eq:Gamma-rf}), the decay rates required to explain MiniBooNE
would correspond to a suppression scale $\Lambda$ of roughly $10^4 \, {\rm TeV}
\lesssim \Lambda \lesssim 10^7 \, {\rm TeV}$. 
If the magnetic moment operator is generated at 1-loop level, we expect
generically
\begin{align}
  \frac{1}{\Lambda} \sim \frac{g}{16\pi^2}\frac{1}{M_{np}} \,,
\end{align}
where $g$ is a coupling constant and $M_{np}$ is the mass scale
of some new physics. We see that for moderately small $g$, $M_{np}$
can be in the TeV range and therefore potentially accessible at the
LHC.

If $N$ is a Majorana neutrino, there will be a contribution to the
light neutrino mass via the type~I seesaw mechanism of order $m_\nu
\simeq m_D^2 / m_N \simeq |U_{\ell 4}|^2 m_N$, where $m_D\simeq
|U_{\ell 4}| m_N$ is the Dirac mass of $N$. In the upper range of the
allowed region for $|U_{\ell 4}|^2$, the seesaw contribution to
$m_\nu$ is too large. However, it is interesting to note that for $m_N
\simeq 250$~MeV and $|U_{\ell 4}|^2 \simeq 10^{-10}$ the seesaw
contribution to $m_\nu$ is of order 0.025~eV, just of the right order
of magnitude for light neutrino masses. Furthermore, the magnetic
moment operator from eq.~\eqref{eq:operator2} will induce also a
contribution to the light neutrino mass via a 1-loop
diagram~\cite{Magill:2018jla}, whose size in general depends on the UV
completion of the operator \eqref{eq:operator2}. Both
contributions---from seesaw and magnetic moment operator---can be
avoided (or suppressed) if $N$ is a Dirac (or pseudo-Dirac)
particle. While our scenario has all ingredients to generate light
neutrino masses, we leave it for future work to identify consistent
models explaining light neutrino masses and mixing in this framework.

\subsection*{Acknowledgments}
We want to thank William Louis for support with respect to technical
aspects of MiniBooNE and associated analyses.  O.F.\ acknowledges
useful discussions with Francois Vanucci, Robert Shrock, and Andreas
Crivellin.
This project is supported by the European Unions Horizon 2020 research
and innovation program under the Marie Sklodowska-Curie grant
agreement No 674896 (Elusives).

\appendix
\section{Heavy neutrino flux at MiniBooNE}\label{sec:flux}

Our starting point is the flux of the muon neutrinos, $\Phi_{\nu_\mu}(p_{\nu_\mu})$, and we focus on the contribution to this flux from kaon decays. These are provided by the MiniBooNE collaboration, cf.\  figs.\ 29 and 31 in ref.\ \cite{AguilarArevalo:2008yp}.
We consider both, neutrino and antineutrino components for each horn polarisation, since it does not matter for the decay signature.

\begin{figure}
\centering
\includegraphics[width=0.5\textwidth]{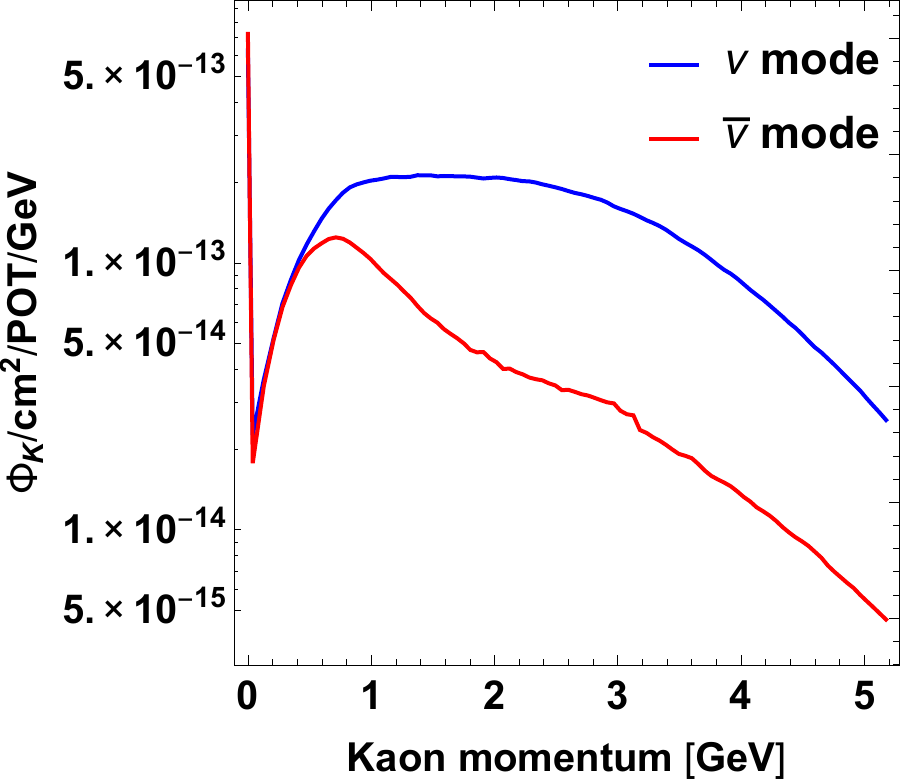}
\caption{Kaon fluxes aiming at the MiniBooNE detector that were obtained from applying an inverse Lorentz boost on the muon (anti) neutrino. For details see text.}
\label{fig:mesonfluxes}
\end{figure}

\paragraph{The kaon flux:} We construct the kaon flux from the light neutrino with the underlying assumptions that for each light neutrino there is one kaon parent and that all of the kaon contribution to the light neutrino flux stems from two body leptonic decays of the kaon (i.e.\ we ignore the three-body decays). An inverse Lorentz transformation allows us to reconstruct the momentum of the kaon $|\vec p_K|$ from the given tables of $|\vec p_\nu|$:
\begin{equation}
|\vec p_K| = \frac{m_K}{2}\left(\frac{|\vec p_\nu|}{p_{\nu_0}} - \frac{p_{\nu_0}}{|\vec p_\nu|} \right)
\label{eq:inverse-LT}
\end{equation}
In the above equation,
\begin{equation}
p_{\nu_0} = \frac{m_K^2 - m_\ell^2}{2 m_K}
\end{equation}
is the definite momentum of a light neutrino from a kaon decay at rest.
Under the above assumptions we can now reconstruct the flux of the parent meson $\Phi_K(p_K)$.
The resulting kaon fluxes, summing $K^+$ and $K^-$ for each of the two horn polarisations, are shown in fig.\ \ref{fig:mesonfluxes}.  The peak for $p_K = 0$ corresponds to stopped kaons which decay at rest.
Notice that our assumptions introduce an error both in shape as well as in magnitude of our prediction, which we take into account in our fit by introducing a 20\% uncorrelated error in each bin.

\paragraph{The heavy neutrino flux:} 
Next we will construct the heavy neutrino flux $\Phi_N(p_N)$.
We start with the assumption that the momenta of the heavy neutrinos are parallel to the parent kaons, i.e.\ $\vec p_N ||\, \vec p_K$. This simplification allows us to construct the heavy neutrino flux $\Phi_N(p_N)$ from the kaon flux $\Phi_K(p_K)$ via Lorentz boosting the momentum $|\vec p_{N,0}|$ from the rest frame of the kaon with momentum $|\vec p_K|$
\begin{equation}
|\vec p_N| = \frac{|\vec p_K| }{ m_K} E_{N,0} + \frac{E_K}{m_K} |\vec p_{N,0}| \cos \theta\,,
\label{eq:Lorentzboost}
\end{equation}
where the heavy neutrino momentum in the meson rest frame is given by
\begin{equation}
|\vec p_{N,0}| = \frac{\sqrt{(m_K^2-\Delta^2)(m_K^2 - \Sigma^2)}}{2 m_K}\,, \qquad \Delta = m_N - m_\ell\,, \qquad \Sigma = m_N + m_\ell\,.
\label{eq:neutrinomomentum}
\end{equation}
For $m_N$ comparable to $m_K$ and for sufficiently large $|\vec p_K|$, also heavy neutrinos that are emitted backwards with respect to $\vec p_K$ can reach the detector.
This means, from the kaon flux we construct two heavy neutrino fluxes: one from the forward emitted $N$ with $\cos \theta = +1$, and one from the backward emitted $N$ with $\cos \theta = -1$.
Both, the backward $\Phi^{\rm bwd}_N(p_N)$ and the forward $\Phi^{\rm fwd}_N(p_N)$ emitted fluxes are normalized to the original light neutrino flux $\Phi_{\nu_\mu} (p_{\nu_\mu})$.
The peak in the kaon spectrum from the stopped kaons (shown in figure \ref{fig:mesonfluxes}) gives rise to monochromatic heavy neutrinos of energy $p_{N,0}$. 
We add this separately to the analysis and call it the ``monochromatic peak''.

\paragraph{Geometrical acceptance:}
We work under the assumption that the kaon momentum is always parallel to the beam line.
In the experiment, neutrinos (light or heavy) are not produced with $\cos \theta = \pm1$, but rather with an angle $\theta = 0+\delta\theta,\pi-\delta\theta$, such that $|\cos\theta| = 1 -\varepsilon$.
This deviation $\varepsilon$ stems from the angles that are smaller than or equal to the solid angle of the detector, which we approximate with $\theta_D \approx \tan \theta_D = r/L$, where
$r$ is the radius of the detector and $L$ is the distance from the source.
The maximal acceptance angle of the heavy neutrino in the lab frame is given by
\begin{eqnarray}
\theta_D = \frac{|p_{N,\bot}|}{|p_{N,\parallel}|} = \frac{p_{N,0} \sin\theta^{\rm rest}_N}{\frac{ p_K }{ m_K} E_{N,0} + \frac{E_K}{m_K}  p_{N,0} \cos \theta^{\rm rest}_N} \, ,
\label{eq:acceptance_angle}
\end{eqnarray}
here $\theta^{\rm rest}_N$ is the kaon rest frame decaying angle. 
The component of the momentum perpendicular to the beam line is not affected by the kaon boost and the parallel one is given by expression \eqref{eq:Lorentzboost}.
For small angles $\sin \theta \sim \theta$, $\cos \theta \sim \pm 1$, the acceptance angle in the rest frame, for the backward and the forward decay, can be easily solved:
\begin{eqnarray}
\theta^{\rm rest}_N = \frac{m_K}{p_{N,0}} \left( p_K  E_{N,0} \pm E_K p_{N,0} \right) \theta_D \, .
\label{eq:acceptance_angle_approximated}
\end{eqnarray}

Since the decay in the rest frame is isotropic, the heavy neutrino flux can be corrected by adding a geometrical factor given by the ratio between the  maximum acceptance angles in the kaon rest frame for the heavy and light neutrinos:
\begin{eqnarray}
f^{\rm fwd} = \frac{\theta^{\rm rest, \, fwd}_{N}}{\theta_\nu^{\rm rest}} \, ; \quad f^{\rm bwd} = \frac{\theta^{\rm rest, \, bwd}_{N}}{\theta_\nu^{\rm rest}} \,.
\label{eq:geometrical_factor}
\end{eqnarray}
Here we assume that the angular acceptance for light neutrinos is already included in $\Phi_{\nu_\mu}$ as provided by the collaboration. For small angles, eq.~\eqref{eq:geometrical_factor} turns into
\begin{eqnarray}
f^{\rm fwd} (p_K) = \frac{ \left( p_K  E_{N,0} + E_K p_{N,0} \right)}{p_{N,0} \left( p_K + E_K \right)} \, ; \quad f^{\rm bwd} (p_K) = \frac{ \left( p_K  E_{N,0} - E_K p_{N,0} \right)}{p_{N,0} \left( p_K + E_K \right)} \, .
\label{eq:geometrical_factor_approximated}
\end{eqnarray}
Note that only the light neutrinos decaying in the forward direction reach the detector, so the heavy neutrino acceptance angle, for both backward and forward directions, have to be compared to the light neutrino one in the forward direction.

We have checked that for the kaon energies at MiniBooNE the small angle approximation \eqref{eq:geometrical_factor_approximated} works very well. On the other hand for the kaon energies in NOMAD, this approximation does not hold because the kaon momentum can be larger.\footnote{Rigorously, equation \eqref{eq:acceptance_angle} has to be solved numerically for both light an heavy neutrinos 
from which one can obtain the ratio of the two kaon rest frame angles.}
We have checked that taking the approximated expression for the geometrical factor for the NOMAD prediction gives an extra enhancement, i.e., we are somewhat over-predicting the number of events in NOMAD. This makes the limit somewhat too strong and is therefore conservative in what concerns the consistency with MiniBooNE, and hence we stick to the approximated expression.

The geometrical factors \eqref{eq:geometrical_factor_approximated} can be expressed as a function of the heavy neutrino momentum performing an inverse boost
\begin{eqnarray*}
\frac{p_K}{m_K} E_N + \frac{E_N}{m_K} p_N = \pm p_{N,0} \, .
\end{eqnarray*}
Solving for $p_K$ we obtained
\begin{eqnarray*}
p_K = \mp \frac{m_K}{m_N^2} E_N p_{N,0} + \frac{m_K}{m_N}\sqrt{ \left( \frac{E_N \, p_{N,0}}{m_N} \right)^2 + p_N^2 - p_{N,0}^2} \, ,
\end{eqnarray*}
where upper (lower) signs apply to the forward (backward) geometrical factors.
Note that in the forward decay case $p_N$ starts from $p_{N,0}$ and in the backward decay from 0.

Finally, the heavy neutrino flux is given by:
\begin{eqnarray*}
\Phi_N(p_N) = f^{\rm fwd} (p_N,m_N) \Phi^{\rm fwd}_N (p_N,m_N) + f^{\rm bwd} (p_N,m_N) \Phi^{\rm bwd}_N (p_N,m_N) \,.
\end{eqnarray*}

\section{Impact of the timing cut}\label{app:timing}

The timing cut of 1.6$\mu$s after each beam spill discussed in
sec.~\ref{sec:timing} has a strong impact on the predicted event
spectrum, since it removes events from ``slow'' heavy neutrinos, which
would provide a less forward peaked angular distribution for the
photon events. In order to illustrate the importance of the timing
cut, we show in this appendix results without requiring arrival within
1.6$\mu$s, i.e., we include all events from $N$ decays in the
predicted signal.

\begin{figure}[t]
    \centering
    \includegraphics[width=0.47\textwidth]{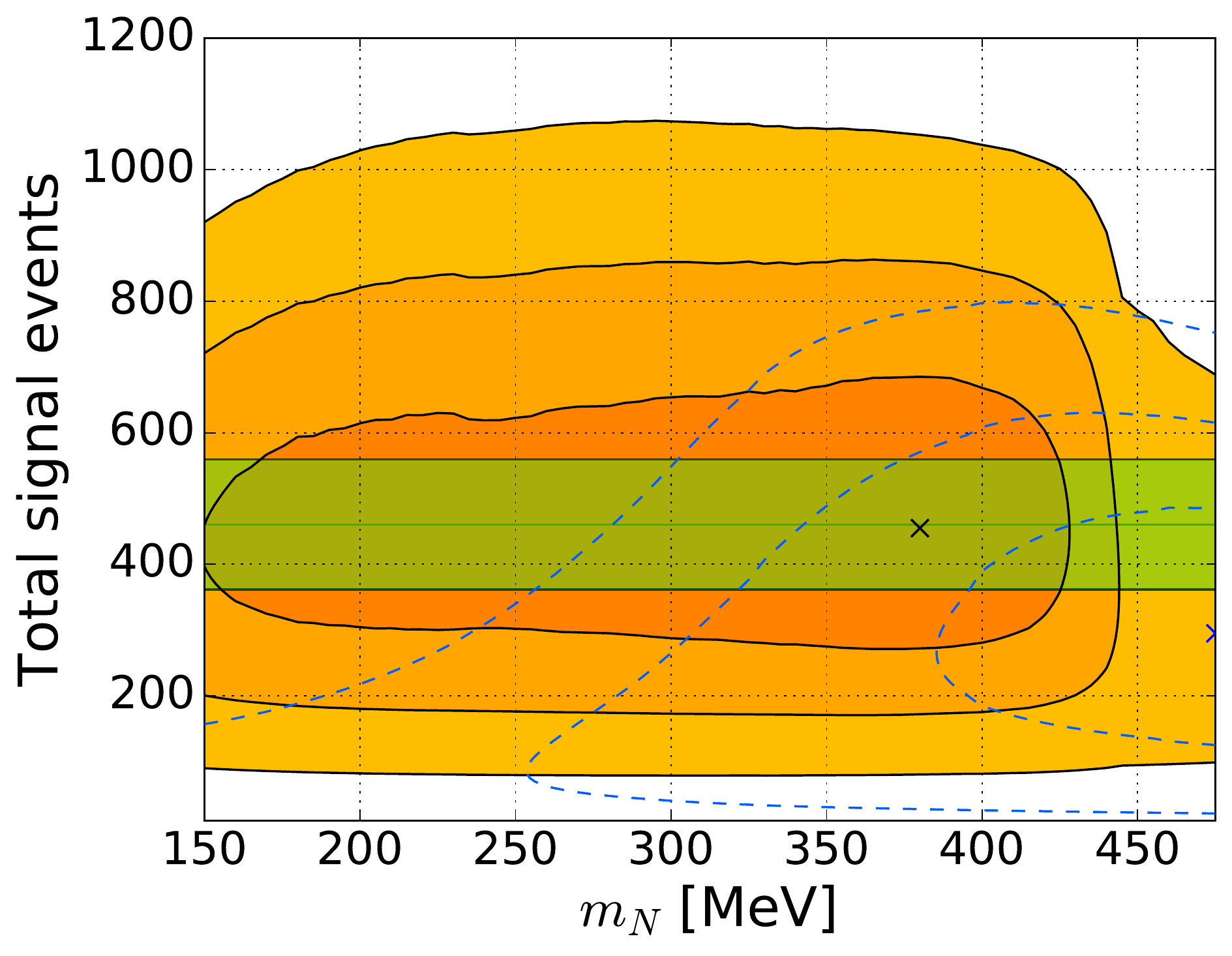}\quad
    \includegraphics[width=0.47\textwidth]{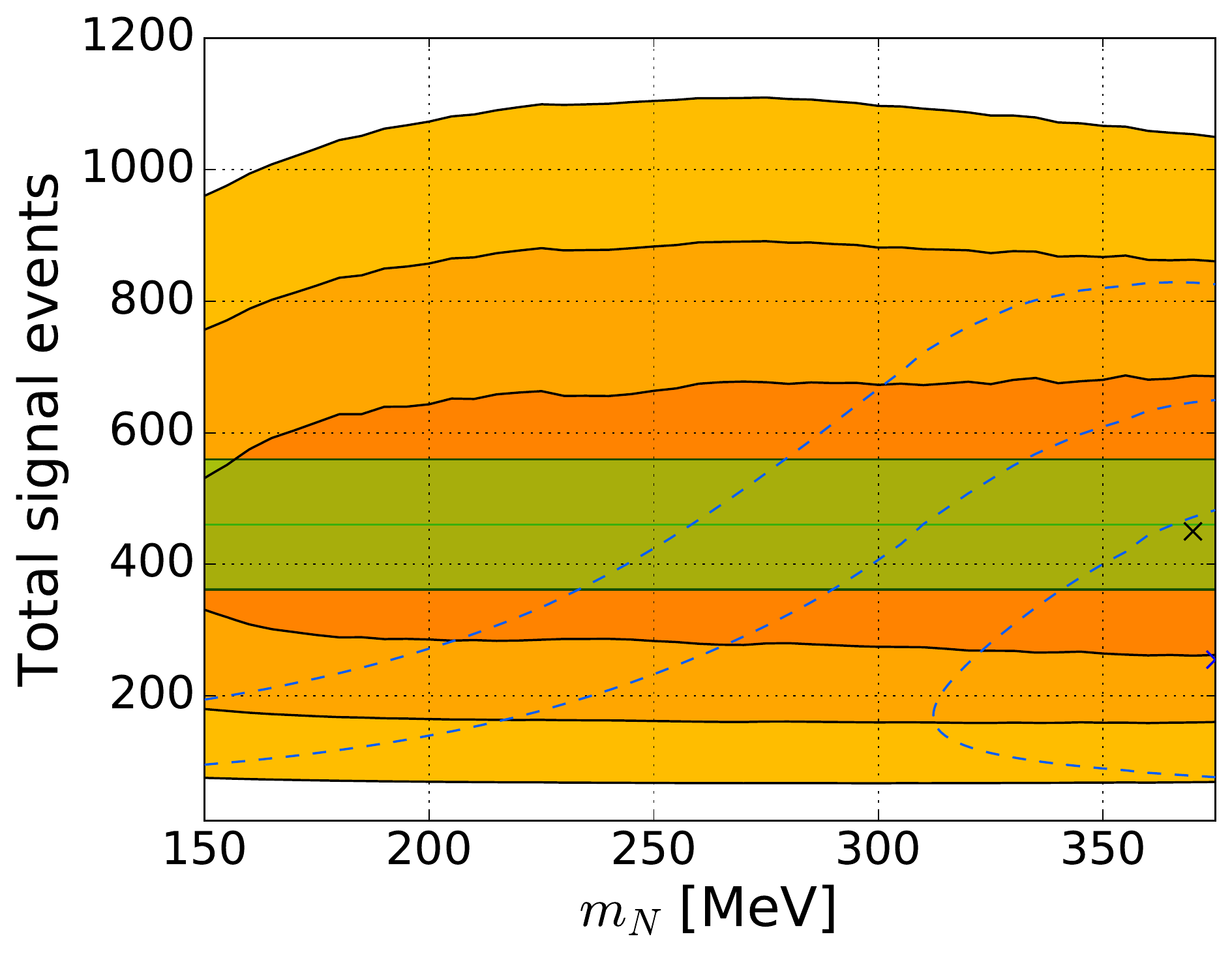}
    \caption{Allowed regions at 1, 2, and 3$\sigma$ for the energy
      (orange regions) and angular (dashed blue curves) in the $N_{\rm
        total}$ versus $m_N$ parameter space, without imposing the
      1.6$\mu$s timing cut. The left (right) panel assumes heavy
      neutrino mixing with the electron (muon). The best fit of the
      energy spectral (angular) fit is indicated with a black (blue)
      cross. In green we show the measured excess of events and its
      1$\sigma$ uncertainty.}
    \label{fig:notiming_2D}
\end{figure}

\begin{figure}
\centering
\includegraphics[width=0.49\linewidth]{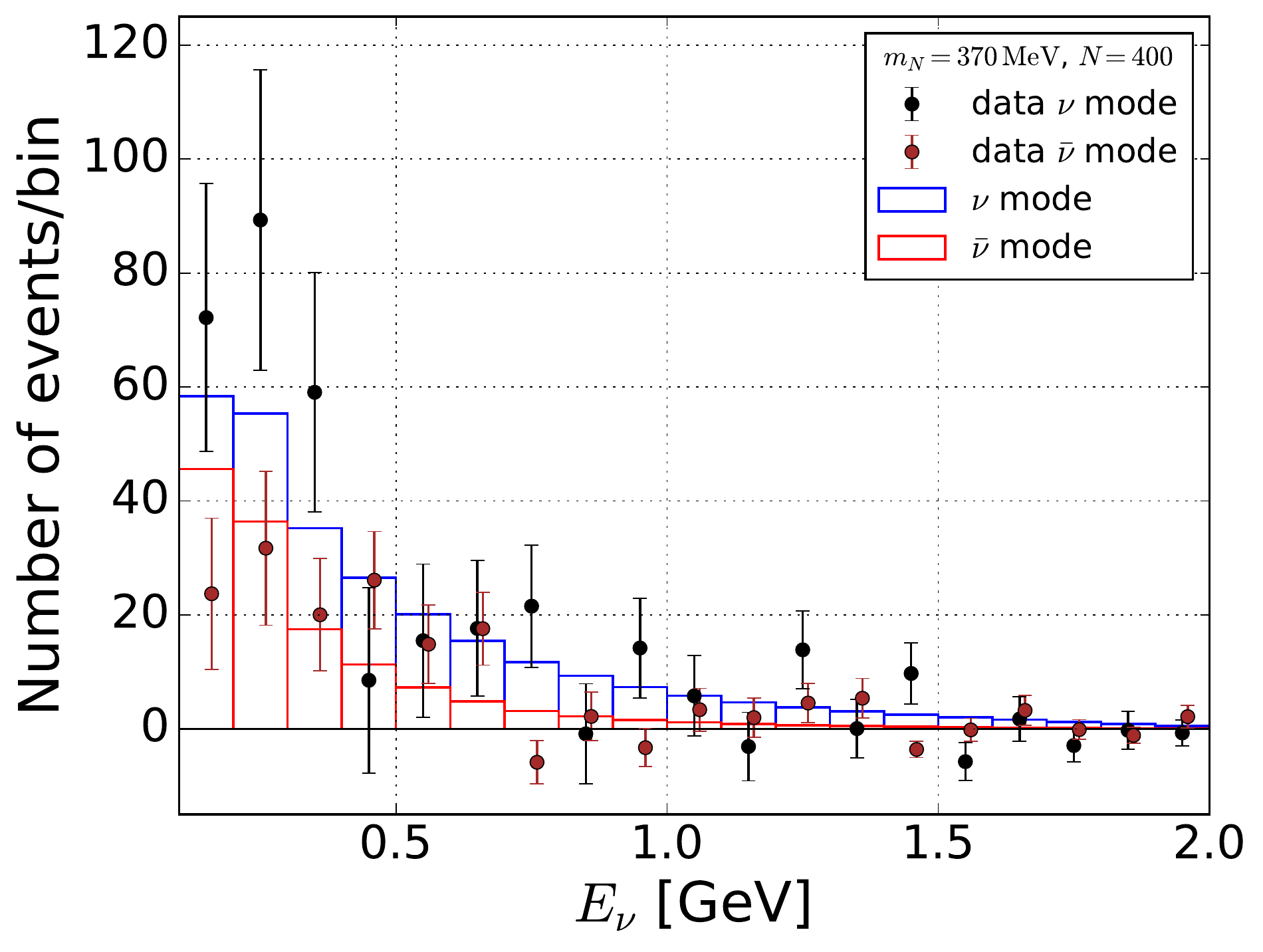}
\hfill
\includegraphics[width=0.49\linewidth]{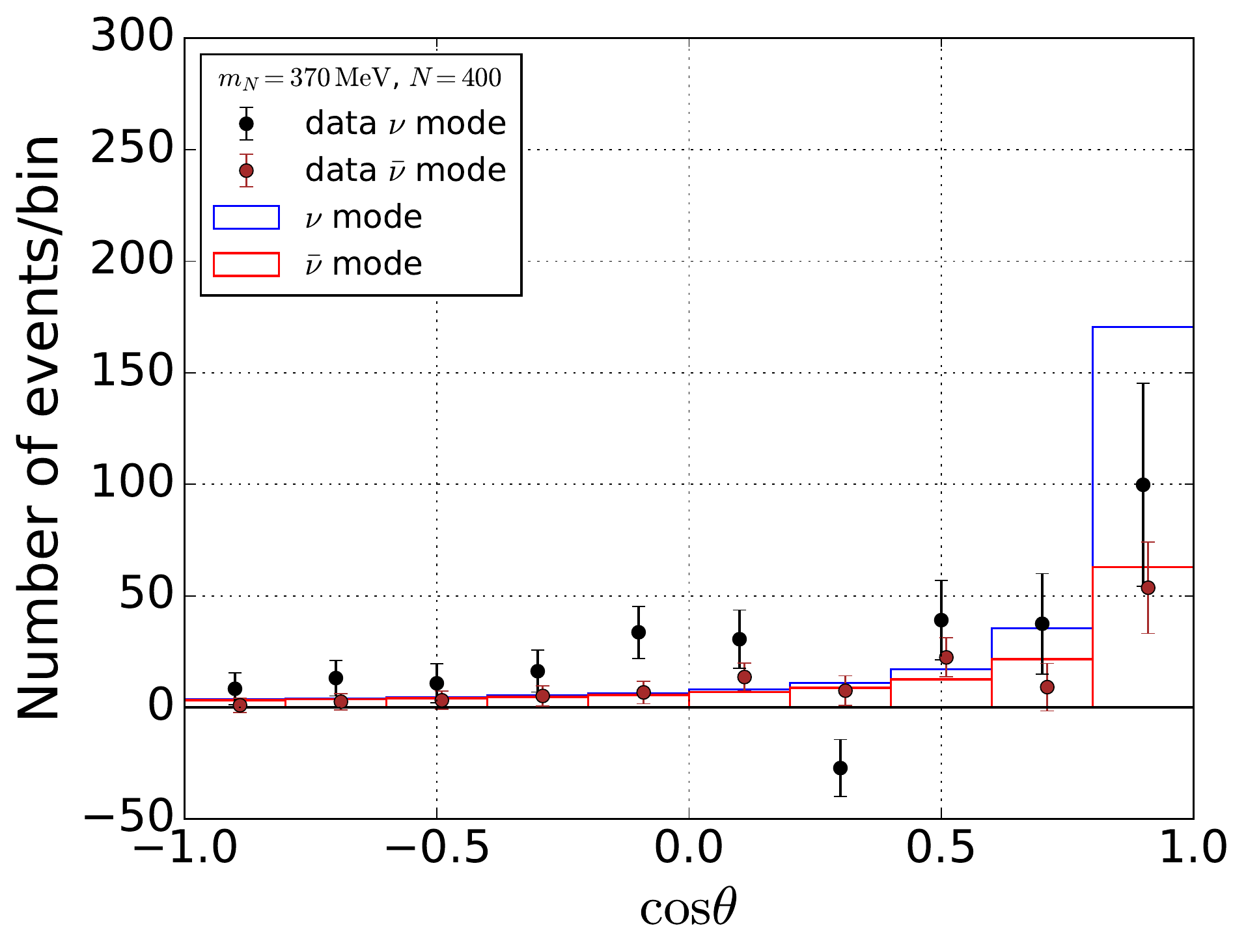}
\caption{Predicted energy (left) and angular (right) spectra for $m_N
  = 370 \, {\rm MeV}$ and $N_{\rm total} = 400$, without imposing the
  1.6$\mu$s timing cut. The dots are the data with background
  subtracted, where each background component is re-scaled by the
  corresponding pull parameter obtained from the fit. Error bars
  include statistical and systematical errors.}
\label{fig:notiming_spectra}
\end{figure} 

From Fig.~\ref{fig:notiming_2D} we see that in this case also the
angular fit shows preference for non-zero signal event numbers, and
the $1\sigma$ allowed regions overlap between the energy and angular
spectral fits. The best fit point for the energy spectrum degrades
only marginally from $\chi^2_{\rm min}/{\rm dof} = 58.1/36$ with
timing cut to 62.8/36 without timing cut in the case of muon
mixing. For the electron mixing we obtain an energy spectrum best fit
with $\chi^2_{\rm min}/{\rm dof} = 61.9/36$. Without the timing cut
the angular fit yields $\chi^2_{\rm min}/{\rm dof} = 32.1/18 \,
(30.0/18)$ for the mixing with the muon (electron), corresponding to a
$p$-value of 1.1\% (3.7\%). In Fig.~\ref{fig:notiming_spectra} we show
the resulting energy and angular spectra for an example point in the
$1\sigma$ overlap region. In comparison with fig.~\ref{energy and
  angular spectra} we clearly observe an improved angular fit, while
still maintaining a good description of the energy distribution. As
discussed in the main text, the formally still rather low $p$-value is
a consequence of the scattered data points with small error bars in
the tail of the distributions.


\hypersetup{linkcolor=black}

\bibliography{refs} 

\providecommand{\href}[2]{#2}\begingroup\raggedright\begin{thebibliography}{10}

\bibitem{Aguilar-Arevalo:2018gpe}
{\bf MiniBooNE}, A.~A. Aguilar-Arevalo et~al.,
  \href{http://dx.doi.org/10.1103/PhysRevLett.121.221801}{{\it {Significant
  Excess of ElectronLike Events in the MiniBooNE Short-Baseline Neutrino
  Experiment}}, } {\em Phys. Rev. Lett.} {\bf 121} (2018), no.~22 221801,
  [\href{http://arxiv.org/abs/1805.12028}{{\tt 1805.12028}}].

\bibitem{Aguilar-Arevalo:2013pmq}
{\bf MiniBooNE}, A.~A. Aguilar-Arevalo et~al.,
  \href{http://dx.doi.org/10.1103/PhysRevLett.110.161801}{{\it {Improved Search
  for $\bar \nu_\mu \rightarrow \bar \nu_e$ Oscillations in the MiniBooNE
  Experiment}}, } {\em Phys. Rev. Lett.} {\bf 110} (2013) 161801,
  [\href{http://arxiv.org/abs/1303.2588}{{\tt 1303.2588}}].

\bibitem{Aguilar:2001ty}
{\bf LSND}, A.~Aguilar-Arevalo et~al.,
  \href{http://dx.doi.org/10.1103/PhysRevD.64.112007}{{\it {Evidence for
  neutrino oscillations from the observation of anti-neutrino(electron)
  appearance in a anti-neutrino(muon) beam}}, } {\em Phys. Rev.} {\bf D64}
  (2001) 112007, [\href{http://arxiv.org/abs/hep-ex/0104049}{{\tt
  hep-ex/0104049}}].

\bibitem{Dentler:2018sju}
M.~Dentler, A.~Hern\'andez-Cabezudo, J.~Kopp, P.~A.~N. Machado, M.~Maltoni,
  et~al., \href{http://dx.doi.org/10.1007/JHEP08(2018)010}{{\it {Updated Global
  Analysis of Neutrino Oscillations in the Presence of eV-Scale Sterile
  Neutrinos}}, } {\em JHEP} {\bf 08} (2018) 010,
  [\href{http://arxiv.org/abs/1803.10661}{{\tt 1803.10661}}].

\bibitem{Gariazzo:2017fdh}
S.~Gariazzo, C.~Giunti, M.~Laveder, and Y.~F. Li,
  \href{http://dx.doi.org/10.1007/JHEP06(2017)135}{{\it {Updated Global 3+1
  Analysis of Short-BaseLine Neutrino Oscillations}}, } {\em JHEP} {\bf 06}
  (2017) 135, [\href{http://arxiv.org/abs/1703.00860}{{\tt 1703.00860}}].

\bibitem{Diaz:2019fwt}
A.~Diaz, C.~A. Arguelles, G.~H. Collin, J.~M. Conrad, and M.~H. Shaevitz, {\it
  {Where Are We With Light Sterile Neutrinos?}},
  \href{http://arxiv.org/abs/1906.00045}{{\tt 1906.00045}}.

\bibitem{Jordan:2018qiy}
J.~R. Jordan, Y.~Kahn, G.~Krnjaic, M.~Moschella, and J.~Spitz,
  \href{http://dx.doi.org/10.1103/PhysRevLett.122.081801}{{\it {Severe
  Constraints on New Physics Explanations of the MiniBooNE Excess}}, } {\em
  Phys. Rev. Lett.} {\bf 122} (2019), no.~8 081801,
  [\href{http://arxiv.org/abs/1810.07185}{{\tt 1810.07185}}].

\bibitem{Ballett:2016opr}
P.~Ballett, S.~Pascoli, and M.~Ross-Lonergan,
  \href{http://dx.doi.org/10.1007/JHEP04(2017)102}{{\it {MeV-scale sterile
  neutrino decays at the Fermilab Short-Baseline Neutrino program}}, } {\em
  JHEP} {\bf 04} (2017) 102, [\href{http://arxiv.org/abs/1610.08512}{{\tt
  1610.08512}}].

\bibitem{Gninenko:2009ks}
S.~N. Gninenko, \href{http://dx.doi.org/10.1103/PhysRevLett.103.241802}{{\it
  {The MiniBooNE anomaly and heavy neutrino decay}}, } {\em Phys. Rev. Lett.}
  {\bf 103} (2009) 241802, [\href{http://arxiv.org/abs/0902.3802}{{\tt
  0902.3802}}].

\bibitem{Gninenko:2010pr}
S.~N. Gninenko, \href{http://dx.doi.org/10.1103/PhysRevD.83.015015}{{\it {A
  resolution of puzzles from the LSND, KARMEN, and MiniBooNE experiments}}, }
  {\em Phys. Rev.} {\bf D83} (2011) 015015,
  [\href{http://arxiv.org/abs/1009.5536}{{\tt 1009.5536}}].

\bibitem{Bertuzzo:2018itn}
E.~Bertuzzo, S.~Jana, P.~A.~N. Machado, and R.~Zukanovich~Funchal,
  \href{http://dx.doi.org/10.1103/PhysRevLett.121.241801}{{\it {Dark Neutrino
  Portal to Explain MiniBooNE excess}}, } {\em Phys. Rev. Lett.} {\bf 121}
  (2018), no.~24 241801, [\href{http://arxiv.org/abs/1807.09877}{{\tt
  1807.09877}}].

\bibitem{Ballett:2018ynz}
P.~Ballett, S.~Pascoli, and M.~Ross-Lonergan,
  \href{http://dx.doi.org/10.1103/PhysRevD.99.071701}{{\it {U(1)' mediated
  decays of heavy sterile neutrinos in MiniBooNE}}, } {\em Phys. Rev.} {\bf
  D99} (2019) 071701, [\href{http://arxiv.org/abs/1808.02915}{{\tt
  1808.02915}}].

\bibitem{Duk:2011yv}
{\bf ISTRA+}, V.~A. Duk et~al.,
  \href{http://dx.doi.org/10.1016/j.physletb.2012.02.087}{{\it {Search for
  Heavy Neutrino in $K^- \to \mu^- \nu_h(\nu_h \to \nu \gamma)$ Decay at ISTRA+
  Setup}}, } {\em Phys. Lett.} {\bf B710} (2012) 307--317,
  [\href{http://arxiv.org/abs/1110.1610}{{\tt 1110.1610}}].

\bibitem{Masip:2012ke}
M.~Masip, P.~Masjuan, and D.~Meloni,
  \href{http://dx.doi.org/10.1007/JHEP01(2013)106}{{\it {Heavy neutrino decays
  at MiniBooNE}}, } {\em JHEP} {\bf 01} (2013) 106,
  [\href{http://arxiv.org/abs/1210.1519}{{\tt 1210.1519}}].

\bibitem{Magill:2018jla}
G.~Magill, R.~Plestid, M.~Pospelov, and Y.-D. Tsai,
  \href{http://dx.doi.org/10.1103/PhysRevD.98.115015}{{\it {Dipole Portal to
  Heavy Neutral Leptons}}, } {\em Phys. Rev.} {\bf D98} (2018), no.~11 115015,
  [\href{http://arxiv.org/abs/1803.03262}{{\tt 1803.03262}}].

\bibitem{Ma:1999im}
E.~Ma, G.~Rajasekaran, and I.~Stancu,
  \href{http://dx.doi.org/10.1103/PhysRevD.61.071302}{{\it {Hierarchical four
  neutrino oscillations with a decay option}}, } {\em Phys. Rev.} {\bf D61}
  (2000) 071302, [\href{http://arxiv.org/abs/hep-ph/9908489}{{\tt
  hep-ph/9908489}}].

\bibitem{PalomaresRuiz:2005vf}
S.~Palomares-Ruiz, S.~Pascoli, and T.~Schwetz,
  \href{http://dx.doi.org/10.1088/1126-6708/2005/09/048}{{\it {Explaining LSND
  by a decaying sterile neutrino}}, } {\em JHEP} {\bf 09} (2005) 048,
  [\href{http://arxiv.org/abs/hep-ph/0505216}{{\tt hep-ph/0505216}}].

\bibitem{Dib:2011hc}
C.~Dib, J.~C. Helo, M.~Hirsch, S.~Kovalenko, and I.~Schmidt,
  \href{http://dx.doi.org/10.1103/PhysRevD.85.011301}{{\it {Heavy Sterile
  Neutrinos in Tau Decays and the MiniBooNE Anomaly}}, } {\em Phys. Rev.} {\bf
  D85} (2012) 011301, [\href{http://arxiv.org/abs/1110.5400}{{\tt 1110.5400}}].

\bibitem{Arguelles:2018mtc}
C.~A. Arguelles, M.~Hostert, and Y.-D. Tsai, {\it {Testing New Physics
  Explanations of MiniBooNE Anomaly at Neutrino Scattering Experiments}},
  \href{http://arxiv.org/abs/1812.08768}{{\tt 1812.08768}}.

\bibitem{Shrock:1980ct}
R.~E. Shrock, \href{http://dx.doi.org/10.1103/PhysRevD.24.1232}{{\it {General
  Theory of Weak Leptonic and Semileptonic Decays. 1. Leptonic Pseudoscalar
  Meson Decays, with Associated Tests For, and Bounds on, Neutrino Masses and
  Lepton Mixing}}, } {\em Phys. Rev.} {\bf D24} (1981) 1232.

\bibitem{Aparici:2009fh}
A.~Aparici, K.~Kim, A.~Santamaria, and J.~Wudka,
  \href{http://dx.doi.org/10.1103/PhysRevD.80.013010}{{\it {Right-handed
  neutrino magnetic moments}}, } {\em Phys. Rev.} {\bf D80} (2009) 013010,
  [\href{http://arxiv.org/abs/0904.3244}{{\tt 0904.3244}}].

\bibitem{Duarte:2015iba}
L.~Duarte, J.~Peressutti, and O.~A. Sampayo,
  \href{http://dx.doi.org/10.1103/PhysRevD.92.093002}{{\it {Majorana neutrino
  decay in an Effective Approach}}, } {\em Phys. Rev.} {\bf D92} (2015), no.~9
  093002, [\href{http://arxiv.org/abs/1508.01588}{{\tt 1508.01588}}].

\bibitem{Butterworth:2019iff}
J.~M. Butterworth, M.~Chala, C.~Englert, M.~Spannowsky, and A.~Titov, {\it
  {Higgs phenomenology as a probe of sterile neutrinos}},
  \href{http://arxiv.org/abs/1909.04665}{{\tt 1909.04665}}.

\bibitem{Coloma:2017ppo}
P.~Coloma, P.~A.~N. Machado, I.~Mart{\'\i ne}z-Soler, and I.~M. Shoemaker,
  \href{http://dx.doi.org/10.1103/PhysRevLett.119.201804}{{\it {Double-Cascade
  Events from New Physics in Icecube}}, } {\em Phys. Rev. Lett.} {\bf 119}
  (2017), no.~20 201804, [\href{http://arxiv.org/abs/1707.08573}{{\tt
  1707.08573}}].

\bibitem{Shoemaker:2018vii}
I.~M. Shoemaker and J.~Wyenberg,
  \href{http://dx.doi.org/10.1103/PhysRevD.99.075010}{{\it {Direct Detection
  Experiments at the Neutrino Dipole Portal Frontier}}, } {\em Phys. Rev.} {\bf
  D99} (2019), no.~7 075010, [\href{http://arxiv.org/abs/1811.12435}{{\tt
  1811.12435}}].

\bibitem{Atre:2009rg}
A.~Atre, T.~Han, S.~Pascoli, and B.~Zhang,
  \href{http://dx.doi.org/10.1088/1126-6708/2009/05/030}{{\it {The Search for
  Heavy Majorana Neutrinos}}, } {\em JHEP} {\bf 05} (2009) 030,
  [\href{http://arxiv.org/abs/0901.3589}{{\tt 0901.3589}}].

\bibitem{Bondarenko:2018ptm}
K.~Bondarenko, A.~Boyarsky, D.~Gorbunov, and O.~Ruchayskiy,
  \href{http://dx.doi.org/10.1007/JHEP11(2018)032}{{\it {Phenomenology of
  GeV-scale Heavy Neutral Leptons}}, } {\em JHEP} {\bf 11} (2018) 032,
  [\href{http://arxiv.org/abs/1805.08567}{{\tt 1805.08567}}].

\bibitem{Dolgov:2000jw}
A.~D. Dolgov, S.~H. Hansen, G.~Raffelt, and D.~V. Semikoz,
  \href{http://dx.doi.org/10.1016/S0550-3213(00)00566-6}{{\it {Heavy Sterile
  Neutrinos: Bounds from Big Bang Nucleosynthesis and Sn1987A}}, } {\em Nucl.
  Phys.} {\bf B590} (2000) 562--574,
  [\href{http://arxiv.org/abs/hep-ph/0008138}{{\tt hep-ph/0008138}}].

\bibitem{Fuller:2009zz}
G.~M. Fuller, A.~Kusenko, and K.~Petraki,
  \href{http://dx.doi.org/10.1016/j.physletb.2008.11.016}{{\it {Heavy Sterile
  Neutrinos and Supernova Explosions}}, } {\em Phys. Lett.} {\bf B670} (2009)
  281--284, [\href{http://arxiv.org/abs/0806.4273}{{\tt 0806.4273}}].

\bibitem{Rembiasz:2018lok}
T.~Rembiasz, M.~Obergaulinger, M.~Masip, M.~A. Perez-Garcia, M.-A. Aloy,
  et~al., \href{http://dx.doi.org/10.1103/PhysRevD.98.103010}{{\it {Heavy
  sterile neutrinos in stellar core-collapse}}, } {\em Phys. Rev.} {\bf D98}
  (2018), no.~10 103010, [\href{http://arxiv.org/abs/1806.03300}{{\tt
  1806.03300}}].

\bibitem{AguilarArevalo:2008yp}
{\bf MiniBooNE}, A.~A. Aguilar-Arevalo et~al.,
  \href{http://dx.doi.org/10.1103/PhysRevD.79.072002}{{\it {The Neutrino Flux
  prediction at MiniBooNE}}, } {\em Phys. Rev.} {\bf D79} (2009) 072002,
  [\href{http://arxiv.org/abs/0806.1449}{{\tt 0806.1449}}].

\bibitem{AguilarArevalo:2008qa}
{\bf MiniBooNE}, A.~A. Aguilar-Arevalo et~al.,
  \href{http://dx.doi.org/10.1016/j.nima.2008.10.028}{{\it {The MiniBooNE
  Detector}}, } {\em Nucl. Instrum. Meth.} {\bf A599} (2009) 28--46,
  [\href{http://arxiv.org/abs/0806.4201}{{\tt 0806.4201}}].

\bibitem{BillLouis}
W.~Louis, {\it private communication},  2019.

\bibitem{CortinaGil:2017mqf}
{\bf NA62}, E.~Cortina~Gil et~al.,
  \href{http://dx.doi.org/10.1016/j.physletb.2018.01.031}{{\it {Search for
  heavy neutral lepton production in $K^+$ decays}}, } {\em Phys. Lett.} {\bf
  B778} (2018) 137--145, [\href{http://arxiv.org/abs/1712.00297}{{\tt
  1712.00297}}].

\bibitem{Artamonov:2014urb}
{\bf E949}, A.~V. Artamonov et~al.,
  \href{http://dx.doi.org/10.1103/PhysRevD.91.059903,
  10.1103/PhysRevD.91.052001}{{\it {Search for heavy neutrinos in
  $K^+\to\mu^+\nu_H$ decays}}, } {\em Phys. Rev.} {\bf D91} (2015), no.~5
  052001, [\href{http://arxiv.org/abs/1411.3963}{{\tt 1411.3963}}]. [Erratum:
  Phys. Rev.D91,no.5,059903(2015)].

\bibitem{Kullenberg:2011rd}
{\bf NOMAD}, C.~T. Kullenberg et~al.,
  \href{http://dx.doi.org/10.1016/j.physletb.2011.11.049}{{\it {A search for
  single photon events in neutrino interactions}}, } {\em Phys. Lett.} {\bf
  B706} (2012) 268--275, [\href{http://arxiv.org/abs/1111.3713}{{\tt
  1111.3713}}].

\bibitem{Vannucci:2014wna}
F.~Vannucci, \href{http://dx.doi.org/10.1155/2014/129694}{{\it {The NOMAD
  Experiment at CERN}}, } {\em Adv. High Energy Phys.} {\bf 2014} (2014)
  129694.

\bibitem{Astier:2003rj}
{\bf NOMAD}, P.~Astier et~al.,
  \href{http://dx.doi.org/10.1016/j.nima.2003.07.054}{{\it {Prediction of
  neutrino fluxes in the NOMAD experiment}}, } {\em Nucl. Instrum. Meth.} {\bf
  A515} (2003) 800--828, [\href{http://arxiv.org/abs/hep-ex/0306022}{{\tt
  hep-ex/0306022}}].

\bibitem{NA62-talk}
E.~Goudzovski, 2019.
\newblock {talk at the International Conference on Kaon Physics 2019, Perugia,
  Italy, 10-13 Sept.\ 2019, https://indico.cern.ch/event/769729}.

\bibitem{Dreiner:2003wh}
H.~K. Dreiner, C.~Hanhart, U.~Langenfeld, and D.~R. Phillips,
  \href{http://dx.doi.org/10.1103/PhysRevD.68.055004}{{\it {Supernovae and
  Light Neutralinos: Sn1987A Bounds on Supersymmetry Revisited}}, } {\em Phys.
  Rev.} {\bf D68} (2003) 055004,
  [\href{http://arxiv.org/abs/hep-ph/0304289}{{\tt hep-ph/0304289}}].

\bibitem{Vannucci:1985vs}
{\bf PS-191}, F.~Vannucci, {\it Decays and oscillations of neutrinos in the
  ps-191 experiment},  in {\em Perspectives in electroweak interactions,
  leptonic session, 20th Rencontres de Moriond, Les Arcs, France, March 17-23,
  1985. Vol. 2}, pp.~277--285, 1985.

\bibitem{Bernardi:1987ek}
G.~Bernardi et~al., \href{http://dx.doi.org/10.1016/0370-2693(88)90563-1}{{\it
  Further limits on heavy neutrino couplings}, } {\em Phys. Lett.} {\bf B203}
  (1988) 332--334.

\bibitem{Bernardi:1986hs}
G.~Bernardi et~al., \href{http://dx.doi.org/10.1016/0370-2693(86)91278-5}{{\it
  {Anomalous Electron Production Observed in the {CERN} Ps Neutrino Beam}}, }
  {\em Phys. Lett.} {\bf B181} (1986) 173--177.

\bibitem{Astier:1989vc}
P.~Astier et~al., \href{http://dx.doi.org/10.1016/0550-3213(90)90516-G}{{\it {A
  Search for Neutrino Oscillations}}, } {\em Nucl. Phys.} {\bf B335} (1990)
  517--545.

\bibitem{Armbruster:2002mp}
{\bf KARMEN}, B.~Armbruster et~al.,
  \href{http://dx.doi.org/10.1103/PhysRevD.65.112001}{{\it Upper limits for
  neutrino oscillations muon-anti-neutrino $\to$ electron-anti-neutrino from
  muon decay at rest}, } {\em Phys. Rev.} {\bf D65} (2002) 112001,
  [\href{http://arxiv.org/abs/hep-ex/0203021}{{\tt hep-ex/0203021}}].

\bibitem{Abe:2019kgx}
{\bf T2K}, K.~Abe et~al., {\it {Search for heavy neutrinos with the T2K near
  detector ND280}},  \href{http://arxiv.org/abs/1902.07598}{{\tt 1902.07598}}.

\bibitem{Abe:2019cer}
{\bf T2K}, K.~Abe et~al.,
  \href{http://dx.doi.org/10.1088/1361-6471/ab227d}{{\it {Search for
  neutral-current induced single photon production at the ND280 near detector
  in T2K}}, } {\em J. Phys.} {\bf G46} (2019), no.~8 08LT01,
  [\href{http://arxiv.org/abs/1902.03848}{{\tt 1902.03848}}].

\bibitem{Antonello:2015lea}
{\bf MicroBooNE, LAr1-ND, ICARUS-WA104}, M.~Antonello et~al., {\it {A Proposal
  for a Three Detector Short-Baseline Neutrino Oscillation Program in the
  Fermilab Booster Neutrino Beam}},
  \href{http://arxiv.org/abs/1503.01520}{{\tt 1503.01520}}.

\bibitem{Machado:2019oxb}
P.~A. Machado, O.~Palamara, and D.~W. Schmitz, {\it {The Short-Baseline
  Neutrino Program at Fermilab}},  {\em Ann. Rev. Nucl. Part. Sci.} {\bf 69}
  (2019) [\href{http://arxiv.org/abs/1903.04608}{{\tt 1903.04608}}].

\bibitem{Alvarez-Ruso:2017hdm}
L.~Alvarez-Ruso and E.~Saul-Sala, {\it {Radiative decay of heavy neutrinos at
  MiniBooNE and MicroBooNE}},  in {\em {Proceedings, Prospects in Neutrino
  Physics (NuPhys2016): London, UK, December 12-14, 2016}}, 2017.
\newblock \href{http://arxiv.org/abs/1705.00353}{{\tt 1705.00353}}.

\end{thebibliography}\endgroup
\bibliographystyle{JHEP_improved}

\end{document}